\documentclass[aps,prd,twocolumn,10pt,groupedaddress]{revtex4-1}
\usepackage{amssymb}
\usepackage{graphicx}
\usepackage{amsmath}
\usepackage{hyperref}
\usepackage{subfigure}
\usepackage{float}
\usepackage{color}
\usepackage{ulem}
\usepackage[utf8]{inputenc}

\begin{document}

\title{Testing weakest force with coldest spot}
\author{Rong-Gen Cai}
\affiliation{CAS Key Laboratory of Theoretical Physics, Institute of Theoretical Physics, Chinese Academy of Sciences, Beijing 100190, China}
\affiliation{School of Fundamental Physics and Mathematical Sciences, Hangzhou Institute for Advanced Study (HIAS), University of Chinese Academy of Sciences, Hangzhou 310024, China}
\author{Shao-Jiang Wang}
\email{schwang@cosmos.phy.tufts.edu (corresponding author)}
\affiliation{CAS Key Laboratory of Theoretical Physics, Institute of Theoretical Physics, Chinese Academy of Sciences, Beijing 100190, China}
\affiliation{Tufts Institute of Cosmology, Department of Physics and Astronomy, Tufts University, 574 Boston Avenue, Medford, Massachusetts 02155, USA}
\affiliation{Quantum Universe Center, Korea Institute for Advanced Study, Seoul 130-722, Korea}
\author{Su Yi}
\author{Jiang-Hao Yu}
\affiliation{CAS Key Laboratory of Theoretical Physics, Institute of Theoretical Physics, Chinese Academy of Sciences, Beijing 100190, China}
\affiliation{School of Physical Sciences, University of Chinese Academy of Sciences, Beijing 100049, China.}

\begin{abstract}
Ultra-cold atom experiment in space with microgravity allows for realization of dilute atomic-gas Bose-Einstein condensate (BEC) with macroscopically large occupation number and significantly long condensate lifetime, which allows for a precise measurement on the shape oscillation frequency by calibrating itself over numerous oscillation periods. In this paper, we propose to measure the Newtonian gravitational constant via ultra-cold atom BEC with shape oscillation, although it is experimentally challenging.  We also make a preliminary perspective on constraining the modified Newtonian potential such as the power-law potential, Yukawa interaction, and fat graviton. A resolution of frequency measurement of $(1-100)\,\mathrm{nHz}$ at most for the occupation number $10^9$, just one order above experimentally  achievable number $N\sim10^6-10^8$, is feasible to constrain the modified  Newtonian potential with Yukawa interaction greatly beyond the current exclusion limits.
\end{abstract}
\maketitle

\section{Introduction}\label{sec:int}

Gravity has been conjectured to be the weakest force among all other fundamental interactions a long time ago \cite{ArkaniHamed:2006dz}, and it is also conjectured to be weaker than the fifth-force mediated by other scalar fields, namely the dubbed scalar weak gravity conjecture \cite{Palti:2017elp} and its strong version \cite{Gonzalo:2019gjp}, from which a refined trans-Planckian censorship conjecture \cite{Cai:2019dzj} is proposed as the bottomline constraint on the effective theories of cosmic inflation. Only until recently, the weak gravity conjecture is elaborated with a general argument from the positivity of black hole entropy shift from higher-dimension operators \cite{Cheung:2018cwt}. Due to the weakestness of gravity, it is generally difficult to constrain deviation beyond the Einstein gravity (or Newtonian gravitational potential in the non-relativistic limit and the  weak gravitational field regime) at small scales \cite{Adelberger:2009zz}, despite of numerous constraints from the astrophysical/cosmological scales \cite{Uzan:2010ri,Jain:2010ka,Baker:2014zba,Berti:2015itd,Sakstein:2015oqa,Koyama:2015vza,Ishak:2018his,Ferreira:2019xrr,Baker:2019gxo}.

To manifest the small-scale effects in experiments, the Bose-Einstein condensate (BEC) is extensively used to collectively exhibit macroscopic quantum  properties that are typically displayed at microscopic scales due to the same quantum identity for all the atoms in a BEC system. However, the cold atom BEC experiments in the terrestrial environment are usually limited by its cooling temperature (nK) and observation time (10 ms - 1 s) due to the large gravity pull on Earth. Therefore, there is a growing interest recently \cite{vanZoest1540,Geiger2011,Becker2018} to implement the ultra-cold atom experiment in space to achieve a colder temperature (pK) and longer observation time (10 s -100 s) thanks to the microgravity environment (gravity gradient around $0.0001\,\mathrm{g/m}$). The ongoing projects have been proposed by NASA, China, ESA, Germany and France, most of which are focusing on the cold atom interferometry to investigate the modified gravity \cite{Tino:2020dsl}, to search for the ultra-light dark matter \cite{Bertoldi:2019tck}, and to detect the gravitational waves \cite{Bertoldi:2019tck} in the sensitivity range between LISA and LIGO. 

In this paper, we will mainly consider the ultra-cold dilute atomic-gas forming a BEC in a spherical harmonic trapping potential with shape oscillation. The long lifetime of BEC state allows for a precise measurement on the shape oscillation frequency, and constraints on the modified gravity is expected in principle. It is worth noting that our preliminary results only serve for theoretical interest since the experimental implements are still quite challenging, which will not be our focus in the current paper.                                                                                                                                                                                                                                                                                          

The paper is organized as follows: In Sec. \ref{sec:GPE}, we derive the general form of the Gross-Pitaevskii equation in the presence of a gravitational potential, from which the frequency deviations in the shape oscillation with respect to the gravity-free case are  derived for some illustrative modified gravity theories in Sec. \ref{sec:theo}. In Sec. \ref{sec:exp}, we give a preliminary perspective on the experimental constraints on these modified gravity theories. We summarize our results in Sec. \ref{sec:con}.

\section{Gross-Pitaevskii equation with gravitational potential}\label{sec:GPE}

Consider $N$ weakly interacting cold atoms of mass $m$ constituting a dilute Bose gas trapped in an external potential $V_\mathrm{ext}$ that would form BEC in the laboratory with aid of laser cooling and evaporative cooling techniques, its mean-field dynamics of the macroscopic wave function $\Psi(\vec{r}_1,\cdots,\vec{r}_N)\approx\Pi_{i=1}^N\psi(\vec{r}_i)$ could be essentially captured by a non-linear Schr\"{o}dinger equation dubbed the Gross-Pitaevskii equation (GPE),
\begin{align}
i\hbar\frac{\partial}{\partial t}\psi(t,\vec{r}\,)&=\left[-\frac{\hbar^2}{2m}\vec{\nabla}^2+V_\mathrm{ext}(\vec{r}\,)\right.\\
&\left.+N\int\mathrm{d}^3\vec{r'}\,\,V_\mathrm{int}(\vec{r}-\vec{r'})|\psi(t,\vec{r'})|^2\right]\psi(t,\vec{r}\,),\nonumber
\end{align}
where the interacting potential $V_\mathrm{int}$ consists of the usual $s$-wave scattering potential as well as a gravitational potential,
\begin{align}
V_\mathrm{int}(\vec{r}-\vec{r'})=g \delta^3(\vec{r}-\vec{r'})+V_G(|\vec{r}-\vec{r'}|)
\end{align}
with $g \equiv4\pi\hbar^2a/m$ characterizing the strength of the $s$-wave scattering of length $a$. Due to the extremely low temperature achieved in the ultra-cold atom experiments, the atoms in BEC are almost collision-less so that all higher-order partial-wave collisions are suppressed at zero collision energy in a short-ranged potential. The external trapping potential $V_\mathrm{ext}$ is assumed to realize a perfect spherically symmetric form in space,
\begin{align}
V_\mathrm{ext}(\vec{r}\,)=\frac12m\omega_0^2r^2,
\end{align}
defining a characteristic length scale for the BEC ground state by
\begin{align}
a_0\equiv\sqrt{\frac{\hbar}{m\omega_0}}.
\end{align}

For the gravitational potential probed by the ultra-cold atom experiment in space, we will focus on the following three kinds of the short-range modifications \cite{Adelberger:2006dh} of the gravitational inverse-square law in the non-relativistic limit and the weak gravitational regime:

\begin{enumerate}
\item Power-law potential:
 \begin{align}\label{eq:power}
 V_G^\mathrm{power}(r)=-\frac{Gm^2}{r}\left[1+\beta_k\left(\frac{1\,\mathrm{mm}}{r}\right)^{k-1}\right],
\end{align}
which could be produced by the simultaneous exchange of multiple massless bosons in the higher-order exchange processes. For example, the $k=2$ case corresponds to the simultaneous exchange of two massless scalar bosons \cite{su:93}; the $k=3$ case corresponds to the simultaneous exchange of massless pseudo-scalar particles between two fermions with the $\gamma_5$-couplings \cite{Ferrer:1998rw}; the $k=5$ case corresponds to the simultaneous exchange of two massless pseudoscalars with the $\gamma_5\gamma^\mu\partial^\mu$ couplings \cite{Ferrer:1998rw} such as the axion or other Goldstone bosons; and the fractional $k$'s are expected from the unparticle exchange \cite{Deshpande:2007mf}.
\item Yukawa interaction:
\begin{align}\label{eq:Yukawa}
V_G^\mathrm{Yukawa}(r)=-\frac{Gm^2}{r}\left[1+\alpha\,\mathrm{e}^{-r/d}\right],
\end{align}
which could be produced by the exchange of natural-parity bosons between unpolarized bodies with the boson mass $\hbar c/d$. 
\item Fat graviton:
\begin{align}
F_\mathrm{fat}(r)&=-\frac{Gm^2}{r^2}\left[1-\exp\left(-\frac{r^3}{\lambda^3}\right)\right],\\
V_G^\mathrm{fat}(r)&=-\frac{Gm^2}{r}\left[1-E_{4/3}\left(\frac{r^3}{\lambda^3}\right)\right]\label{eq:fat}
\end{align}
with the exponential integral function $E_\nu(z)=\int_1^\infty\mathrm{e}^{-zt}t^{-\nu}\mathrm{d}t$. Here the graviton is conjectured to be a ``fat'' object with size $\lambda$ \cite{Sundrum:2003jq}, and the gravitational force falls off rapidly to zero at $r\to0$  limit  \cite{Adelberger:2006dh}.
\end{enumerate}

\subsection{Equation of motion}\label{subsec:EOM}

The GPE could be derived from the variation of the action \cite{PhysRevLett.77.5320,PhysRevA.56.1424,Gupta:2015cta}
\begin{align}
S&=\int\mathrm{d}tL=\int\mathrm{d}t\int\mathrm{d}^3\vec{r}\,\,\mathcal{L},\\
\mathcal{L}&=\frac{i\hbar}{2}\left(\psi\frac{\partial\psi^*}{\partial t}-\psi^*\frac{\partial\psi}{\partial t}\right)+\frac{\hbar^2}{2m}\vec{\nabla}\psi^*\cdot\vec{\nabla}\psi+V_\mathrm{ext}|\psi|^2\nonumber\\
&+\frac{g  N}{2}|\psi|^4+\frac{N}{2}|\psi|^2\int\mathrm{d}^3\vec{r'}\,\,V_G(|\vec{r}-\vec{r'}|)|\psi(t,\vec{r'})|^2
\end{align}
with respect to $\delta\psi$ and $\delta\psi^*$. For a normalized wave function ansatz,
\begin{align}
\psi(t,\vec{r}\,)=\frac{\mathrm{e}^{i\left[\gamma(t)+B(t)r^2\right]}}{(\sqrt{\pi}\sigma(t))^{3/2}}\mathrm{e}^{-\frac{r^2}{2\sigma(t)^2}},
\end{align}
the width parameter $\sigma(t)$ and phase parameters $\gamma(t)$ and $B(t)$ could be solved from the corresponding Euler-Lagrange equations of Lagrangian
\begin{align}\label{eq:L}
L&=\int\mathrm{d}^3\vec{r}\,\,\mathcal{L}=\hbar\dot{\gamma}+L_G+\frac{g  N}{4\sqrt{2\pi^3}\sigma^3}\nonumber\\
&+\frac32\sigma^2\left(\hbar\dot{B}+\frac{2\hbar^2}{m}B^2+\frac{\hbar^2}{2m\sigma^4}+\frac12m\omega_0^2\right),
\end{align}
where 
\begin{align}\label{eq:LG}
L_G&=\frac{N}{2}\int\mathrm{d}^3\vec{r}_1\,|\psi(t,\vec{r}_1)|^2\int\mathrm{d}^2\vec{r}_2\,V_G(|\vec{r}_1-\vec{r}_2|)|\psi(t,\vec{r}_2)|^2\nonumber\\
&=\frac{N}{2\pi^3}\int\mathrm{d}^3\vec{r}_1\mathrm{d}^3\vec{r}_2V_G(|\vec{r}_1-\vec{r}_2|)\frac{1}{\sigma^6}\mathrm{e}^{-\frac{r_1^2+r_2^2}{\sigma^2}}
\end{align}
will be analytically evaluated later with specific form of the gravitational potential.

The total time-derivative term $\hbar\dot{\gamma}(t)$ in \eqref{eq:L} does not admit a dynamical equation, and the rest of Euler-Lagrange equations for $\sigma(t)$ and $B(t)$ read
\begin{align}
B(t)=\frac{m\dot{\sigma}(t)}{2\hbar\sigma(t)},
\end{align}
\begin{align}
\hbar\dot{B}+\frac{2\hbar^2}{m}B^2-\frac{\hbar^2}{2m\sigma^4}+\frac12m\omega_0^2=\frac{g  N}{4\sqrt{2\pi^3}\sigma^5}-\frac{1}{3\sigma}\frac{\mathrm{d}L_G}{\mathrm{d}\sigma},
\end{align}
respectively, which could be combined into a single equation of motion (EOM)
\begin{align}\label{eq:EOMsigma}
m\ddot{\sigma}=-m\omega_0^2\sigma+\frac{\hbar^2}{m\sigma^3}+\frac{g  N}{2\sqrt{2\pi^3}\sigma^4}-\frac23\frac{\mathrm{d}L_G}{\mathrm{d}\sigma}.
\end{align}
After adopting following dimensionless quantities,
\begin{align}
\tau=\omega_0t, \quad \nu=\frac{\sigma}{a_0},
\end{align}
the EOM \eqref{eq:EOMsigma} becomes
\begin{align}\label{eq:EOMnu}
\frac{\mathrm{d}^2\nu}{\mathrm{d}\tau^2}+\frac{\mathrm{d}U}{\mathrm{d}\nu}=0,
\end{align}
which effectively describes a particle moving in a dimensionless effective potential
\begin{align}\label{eq:Ueff}
U(\nu)=\frac12\left(\nu^2+\frac{1}{\nu^2}\right)+\sqrt{\frac{2}{\pi}}\frac{N}{3\nu^3}\frac{a}{a_0}+\frac23\frac{L_G}{\hbar\omega_0}.
\end{align}

\begin{figure}
\centering
\includegraphics[width=0.49\textwidth]{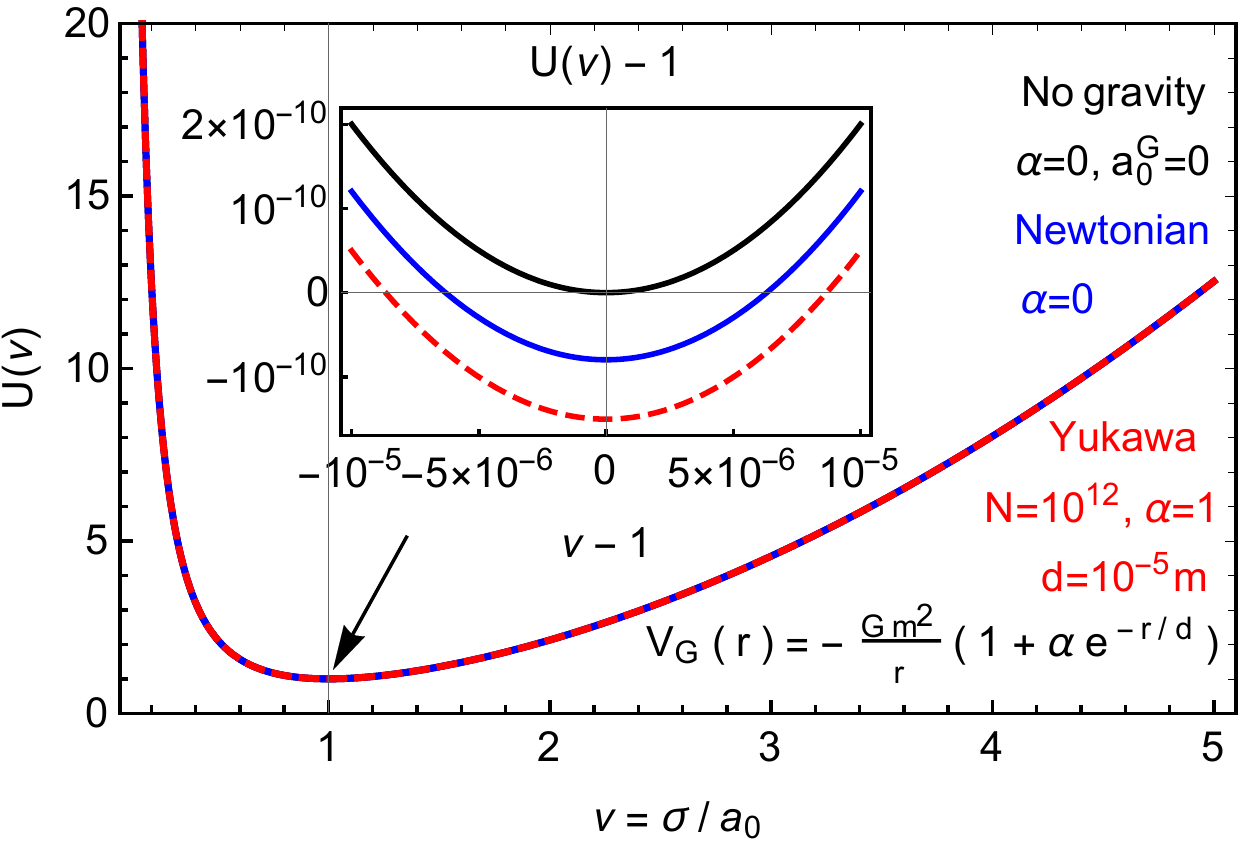}
\caption{The effective potential for BEC oscillation in the ultra-cold atom experiment with magnetically manipulated scattering length $a=0$. The gravitational potentials are illustrated for the cases without gravity (black solid line), with Newtonian gravity only (blue solid line), and with Yukawa interaction (red dashed line). }\label{fig:Ueff}
\end{figure}

To evaluate $L_G$ in  \eqref{eq:Ueff}, we first make replacements of variables as
\begin{align}
\vec{r}=\frac{1}{\sqrt{2}}\left(\vec{r}_1-\vec{r}_2\right), \quad \vec{R}=\frac{1}{\sqrt{2}}\left(\vec{r}_1+\vec{r}_2\right),
\end{align}
so that $r_1^2+r_2^2=r^2+R^2$, and then $L_G$ becomes
\begin{align}
L_G=\frac{N}{2\pi^3\sigma^6}\int_0^\infty\mathrm{d}R(4\pi R^2)\mathrm{e}^{-\frac{R^2}{\sigma^2}}\int_0^\infty\mathrm{d}r(4\pi r^2)V_G(\sqrt{2}r)\mathrm{e}^{-\frac{r^2}{\sigma^2}}.
\end{align}
After employing the following dimensionless quantities
\begin{align}
x=\frac{r}{a_0}, \quad a_0^G=\frac{Gm^2}{\hbar\omega_0}\equiv\frac{c}{\omega_0}\frac{m^2}{m_\mathrm{Pl}^2},
\end{align}
with the Planck mass $m_\mathrm{Pl}\equiv\sqrt{\hbar c/G}$, one arrives at
\begin{align}
\frac{L_G}{\hbar\omega_0}=\frac{N}{2\pi^{3/2}}\frac{a_0^G}{a_0}\frac{I(\nu)}{\nu^3},
\end{align}
where the integration $I(\nu)$ reads
\begin{align}\label{eq:integral}
I(\nu)=\int_0^\infty\mathrm{d}x(4\pi x^2)\left[\frac{a_0}{Gm^2}V_G(\sqrt{2}a_0x)\right]\mathrm{e}^{-\frac{x^2}{\nu^2}}.
\end{align}
Therefore, the EOM could be solved for given effective potential of form
\begin{align}\label{eq:UeffVG}
U(\nu)=\frac12\left(\nu^2+\frac{1}{\nu^2}\right)+\sqrt{\frac{2}{\pi}}\frac{N}{3\nu^3}\frac{a}{a_0}+\frac{N}{3\pi^{3/2}}\frac{a_0^G}{a_0}\frac{I(\nu)}{\nu^3}.
\end{align}
If $U(\nu)$ admits a local minimum $\nu_\mathrm{min}$, the width of BEC  $\sigma(t)=\nu(t)a_0$ would experience the shape oscillation with frequency $\omega=\sqrt{U''(\nu_\mathrm{min})}\omega_0$ around $\nu_\mathrm{min}$. An illustration for the effective potential is shown in Fig. \ref{fig:Ueff} with magnetically manipulated scattering length $a=0$ (as we will assume later), and the cases without gravity (black solid line), with the Newtonian gravity only (blue solid line), and with extra Yukawa interaction (red dashed line), which will be presented in detail in the following sections in addition to the cases with the extra power-law potential and the fat-graviton potential.

\subsection{Shape oscillation frequency}\label{subsec:resonance}

To estimate the contributions of each term in the effective potential \eqref{eq:UeffVG}, one could adopt the typical value of $a_0\sim10^{-4}\,\mathrm{cm}$ and $N\sim10^6-10^8$, which is experimentally achievable currently, as well as $a_0^G$ taken to be
\begin{align}
a_0^G=\frac{c}{\omega_0}\left(\frac{m}{m_\mathrm{Pl}}\right)^2=1.747\times10^{-28}\,\mathrm{cm}\left(\frac{\omega_0}{\mathrm{Hz}}\right)^{-1}A^2,
\end{align}
where the atomic mass number $A$ is usually around order of hundred, and the trapping frequency $\omega_0$ could be adjusted in the range of $50-10,000$ Hz. Therefore, for typical choice of scattering length $a\approx a_0\gg a_0^G$ due to the $m/m_\mathrm{Pl}$ suppression in the ratio $a_0^G/a_0\sim10^{-24}-10^{-22}$, the second term would dominate over the third term in \eqref{eq:UeffVG}, namely, the $s$-wave scattering would totally erase the trace of gravitational potential. To manifest the gravitational effect, one could tune down the scattering length $a$ magnetically to zero \cite{Cornish:2000zz} (Hereafter we will assume this idealistic condition for our preliminary studies) so that we can get rid of the second term in \eqref{eq:UeffVG}, and the effect of additional gravitational potential could be extracted from the frequency deviation of shape oscillation \cite{PhysRevA.47.4114,Inouye1998,PhysRevLett.81.69,PhysRevLett.81.5109} in the absence of gravity.

\subsubsection{Without gravity}

In the case without gravity \cite{PhysRevA.56.1424}, the effective potential reads
\begin{align}
U(\nu)=\frac12\left(\nu^2+\frac{1}{\nu^2}\right),
\end{align}
and the corresponding EOM
\begin{align}
\frac{\mathrm{d}^2\nu}{\mathrm{d}\tau^2}+\nu-\frac{1}{\nu^3}=0
\end{align}
is solved by
\begin{align}
\nu_0(\tau)=\frac{1}{\nu_i}\sqrt{\frac{1+\nu_i^4+(\nu_i^4-1)\cos2\tau}{2}}
\end{align}
with initial condition $\nu(\tau=0)=\nu_i$. The corresponding local minimum $\nu_\mathrm{min}=1$ could be solved from the zeros of the effective potential $U'(\nu)=0$, around which the oscillation frequency is exactly $\omega(\nu_\mathrm{min})\equiv\sqrt{U''(\nu_\mathrm{min})}\omega_0=2\omega_0$ to the second order. Higher order terms in the expansion of $\nu(\tau)$ around $\nu_\mathrm{min}$, which should be extracted by recording the full time evolution of shape oscillation in the experiment, could be ignored in principle by controlling the initial size $\nu_i$ near $\nu_\mathrm{min}$.

\subsubsection{With Newtonian gravity only}

As for the shape oscillation with only Newtonian gravity included, the effective potential is computed as
\begin{align}\label{eq:NewtonUeff}
U(\nu)=\frac12\left(\nu^2+\frac{1}{\nu^2}\right)-\sqrt{\frac{2}{\pi}}\frac{Na_0^G}{3a_0\nu},
\end{align}
from which the oscillation frequency is found to be deviated from $2\omega_0$ by
\begin{align}\label{eq:Newtonomega}
\frac{\omega}{2\omega_0}=\sqrt{1+\frac14\left(\frac{1}{\nu_\mathrm{min}^4}-1\right)},
\end{align}
where the local minimum $\nu_\mathrm{min}$ is the root of $U'(\nu)=0$, namely,
\begin{align}\label{eq:Newtonnumin}
\frac{1}{\nu_\mathrm{min}^4}-1=\sqrt{\frac{2}{\pi}}\frac{Na_0^G}{3a_0\nu_\mathrm{min}^3}.
\end{align}
Due to the suppression factor $a_0^G/a_0\ll1$, deviations on both the local minimum $\nu_\mathrm{min}$ from 1 and oscillation frequency $\omega$ from $2\omega_0$, respectively, are extremely small. 

Nevertheless, if the oscillation frequency could be measured to a  high accuracy, we are able to obtain the Newtonian gravitational constant in this BEC experiment. In fact, the combination of \eqref{eq:Newtonomega} and \eqref{eq:Newtonnumin} gives rise to an expression for the Newtonian gravitational constant as
\begin{align}\label{eq:GN}
G=\sqrt{\frac{\pi}{2}}\frac{3\sigma_\mathrm{min}^3}{mN}(\omega^2-4\omega_0^2),
\end{align}
from which the derivative with respect to frequency change reads
\begin{align}\label{eq:dGNdv}
\frac{\mathrm{d}G}{\mathrm{d}\omega}=\frac{3G}{\sigma_\mathrm{min}}\frac{\mathrm{d}\sigma_\mathrm{min}}{\mathrm{d}\omega}+3\sqrt{2\pi}\frac{\sigma_\mathrm{min}^3}{mN}\omega.
\end{align}
To further evaluate the part of $\mathrm{d}\sigma_\mathrm{min}/\mathrm{d}\omega$, one first rewrites  \eqref{eq:Newtonnumin} as
\begin{align}
a_0^4=\sigma_\mathrm{min}^4+\sqrt{\frac{2}{\pi}}\frac{mN}{3\omega_0^2}G\sigma_\mathrm{min},
\end{align}
which, after taking derivative, becomes
\begin{align}
0=4\sigma_\mathrm{min}^3\frac{\mathrm{d}\sigma_\mathrm{min}}{\mathrm{d}\omega}+\sqrt{\frac{2}{\pi}}\frac{mN}{3\omega_0^2}\frac{\mathrm{d}(G\sigma_\mathrm{min})}{\mathrm{d}\omega}.
\end{align}
Here the derivative term $\mathrm{d}(G\sigma_\mathrm{min})/\mathrm{d}\omega$ could be replaced by taking the derivative of \eqref{eq:GN} after multiplied with $\sigma_\mathrm{min}$, namely,
\begin{align}
\frac{\mathrm{d}(G\sigma_\mathrm{min})}{\mathrm{d}\omega}=3\sqrt{2\pi}\frac{a_0^4}{mN}\frac{\omega\omega_0^4}{(\omega^2-3\omega_0^2)^2}.
\end{align}
Then $\mathrm{d}\sigma_\mathrm{min}/\mathrm{d}\omega$ could be  obtained from above two equations as
\begin{align}
\frac{\mathrm{d}\sigma_\mathrm{min}}{\mathrm{d}\omega}=-\frac{\omega\sigma_\mathrm{min}^5}{2\omega_0^2a_0^4}.
\end{align}
Now \eqref{eq:dGNdv} could be exactly calculated as
\begin{align}
\frac{1}{G}\frac{\mathrm{d}G}{\mathrm{d}\omega}=\frac{\omega^3}{2(\omega^2-4\omega_0^2)(\omega^2-3\omega_0^2)}
\end{align} 
without explicitly finding the root of \eqref{eq:Newtonnumin}. Finally, by approximating $\omega\approx2\omega_0$ in the factors $\omega+2\omega_0$ and $\omega^2-3\omega_0^2$, we arrive at a preliminary estimation for the relative error of $G$ from the relative error of $\omega$ as
\begin{align}\label{eq:dGbyG}
\frac{\Delta G}{G}\approx\frac{\Delta\omega}{\omega}\bigg/\left(\frac{\omega}{2\omega_0}-1\right),
\end{align}
which is suppressed by the frequency deviation $\omega/(2\omega_0)-1$. 

As we will see later at the Newtonian limit of modified gravity theories, the frequency deviation $\omega/(2\omega_0)-1$ is usually of the size $10^{-14}$ for $N=10^9$ just above the current achievable $N\sim10^6-10^8$, which requires the relative error on the $\omega$ as small as $10^{-18}$ if we want to measure $\Delta G/G$ up to precision of  $10^{-4}$. This corresponds to a resolution of frequency measurement $\Delta\omega\sim10^{-18}\omega\sim10^{-16}-10^{-14}\,\mathrm{Hz}$ provided that the value of $\omega_0$ ranges from $50-10000\,\mathrm{Hz}$. However, the practical resolution of frequency measurement could be relaxed since the oscillation frequency could be measured and calibrated over numerous oscillation periods for sufficiently long lifetime of the BEC state like that in space.  To our knowledge, despite of the atom interferometric determination of the Newtonian gravitational constant \cite{Rosi:2014kva,Fixler:2007is},  our proposal for the measurement on Newtonian gravitational constant  from the cold atom BEC experiment with shape oscillation has not been investigated  in the literature, although it is experimentally more challenging.

On the other hand, with Newtonian gravitational constant known from other experiments, we can further probe the realms of modified gravity theories by measuring the deviation of the oscillation frequency. Note that the impact on the relative error of the oscillation frequency due to the Newtonian gravitational potential is much smaller than the frequency deviation as seen from \eqref{eq:dGbyG}, namely,
\begin{align}\label{eq:dwbyw}
\frac{\Delta\omega}{\omega}\approx\frac{\Delta G}{G}\left(\frac{\omega}{2\omega_0}-1\right)\ll\left(\frac{\omega}{2\omega_0}-1\right),
\end{align}
as long as the frequency deviation is small, $\omega\approx2\omega_0$, and the Newtonian gravitational constant could be measured from other experiments to relatively high accuracy, $\Delta G/G\ll1$.

\section{Shape oscillation for modified gravitational potential}\label{sec:theo}

\subsection{Power-law potential}\label{subsec:power}

For an additional power-law potential \eqref{eq:power} to the Newtonian gravity, the corresponding effective potential is obtained as
\begin{align}
U(\nu)&=\frac12\left(\nu^2+\frac{1}{\nu^2}\right)\\
&-\sqrt{\frac{2}{\pi}}\frac{Na_0^G}{3a_0\nu}\left[1+\beta_k\left(\frac{1\,\mathrm{mm}/a_0}{\sqrt{2}\nu}\right)^{k-1}\Gamma\left(\frac{3-k}{2}\right)\right]\nonumber
\end{align}
for $k<3$. For $k\geq3$, the integral \eqref{eq:integral} is not well-defined. Similar to \eqref{eq:Newtonomega}, the oscillation frequency could be found to be deviated from $2\omega_0$ by
\begin{align}\label{eq:omegapower}
\frac{\omega}{2\omega_0}&=\sqrt{1+\frac{2-k}{4}\left(\frac{1}{\nu_\mathrm{min}^4}-1\right)+\frac{k-1}{4}\sqrt{\frac{2}{\pi}}\frac{Na_0^G}{3a_0\nu_\mathrm{min}^3}},
\end{align}
where the local minimum $\nu_\mathrm{min}$ is determined by $U'(\nu)=0$ as
\begin{align}
\frac{1}{\nu_\mathrm{min}^4}-1=&\sqrt{\frac{2}{\pi}}\frac{Na_0^G}{3a_0\nu_\mathrm{min}^3}\left[1+\bigg.\right.\nonumber\\
&\left.k\beta_k\left(\frac{1\,\mathrm{mm}/a_0}{\sqrt{2}\nu_\mathrm{min}}\right)^{k-1}\Gamma\left(\frac{3-k}{2}\right)\right].
\end{align}

\subsection{Yukawa interaction}\label{subsec:Yukawa}

For an additional Yukawa interaction \eqref{eq:Yukawa} to the Newtonian gravity, the corresponding effective potential is obtained as
\begin{align}
U(\nu)&=\frac12\left(\nu^2+\frac{1}{\nu^2}\right)-\sqrt{\frac{2}{\pi}}\frac{Na_0^G}{3a_0\nu}\nonumber\\
&\times\left\{1+\alpha\left[1-\sqrt{\pi}\left(\frac{\beta\nu}{\sqrt{2}}\right)\mathrm{e}^{\frac{\beta^2\nu^2}{2}}\mathrm{Erfc}\left(\frac{\beta\nu}{\sqrt{2}}\right)\right]\right\},
\end{align}
where $\beta\equiv a_0/d$ and $\mathrm{Erfc}(z)\equiv1-\mathrm{Erf}(z)$ is the complementary error function. The oscillation frequency could be found to be deviated from $2\omega_0$ by
\begin{align}\label{eq:omegaYukawa}
\frac{\omega}{2\omega_0}=&\left\{1+\left(\frac{1}{\nu_\mathrm{min}^4}-1\right)\left(1+\frac{\beta^2\nu_\mathrm{min}^2}{4}\right)\right.\nonumber\\
&\left.-\sqrt{\frac{2}{\pi}}\frac{Na_0^G}{a_0}\frac{3+3\alpha+\beta^2v_\mathrm{min}^2}{12v_\mathrm{min}^3}\right\}^{1/2},
\end{align}
where the local minimum $\nu_\mathrm{min}$ is determined by $U'(\nu)=0$ as
\begin{align}
\frac{1}{\nu_\mathrm{min}^4}-1=&\sqrt{\frac{2}{\pi}}\frac{Na_0^G}{3a_0\nu_\mathrm{min}^3}\left[1+\alpha-\alpha\beta^2\nu_\mathrm{min}^2\right.\nonumber\\
&\left.+\sqrt{\frac{\pi}{2}}\alpha\beta^3\nu_\mathrm{min}^3\mathrm{Erfc}\left(\frac{\beta\nu_\mathrm{min}}{\sqrt{2}}\right)\right].
\end{align}

\subsection{Fat graviton}\label{subsec:fat}

For an additional fat-graviton potential \eqref{eq:fat} to the Newtonian gravity, the corresponding effective potential is obtained as
\begin{align}
U(\nu)&=\frac12\left(\nu^2+\frac{1}{\nu^2}\right)-\sqrt{\frac{2}{\pi}}\frac{Na_0^G}{3a_0\nu}\times\nonumber\\
&\left[1-\frac{1}{2\beta^2\nu^2}\Gamma\left(\frac53\right)\,_2F_2\left(\frac12,\frac56;\frac23,\frac32;-\frac{1}{54\beta^6\nu^6}\right)\right.\nonumber\\
&-\frac{1}{45\beta^4\nu^4}\Gamma\left(-\frac23\right)\,_2F_2\left(\frac56,\frac76;\frac43,\frac{11}{6};-\frac{1}{54\beta^6\nu^6}\right)\nonumber\\
&\left.-\frac{1}{56\beta^6\nu^6}\,_3F_3\left(1,\frac76,\frac32;\frac43,\frac53,\frac{13}{6};-\frac{1}{56\beta^6\nu^6}\right)\right],
\end{align}
where $\beta\equiv a_0/\lambda$ and $\,_pF_q(a_1,\cdots,a_p;b_1,\cdots,b_q;z)$ is the generalized hypergeometric function. The oscillation frequency could be found to be deviated from $2\omega_0$ by
\begin{align}\label{eq:omegafat}
\frac{\omega}{2\omega_0}&=\left\{1+\frac34\left(\frac{1}{\nu_\mathrm{min}^4}-1\right)-\sqrt{\frac{2}{\pi}}\frac{Na_0^G}{6a_0\nu_\mathrm{min}^3}\left[1\bigg.\right.\right.\nonumber\\
&-\frac{3}{\beta^2\nu_\mathrm{min}^2}\Gamma\left(\frac53\right)\,_1F_1\left(\frac56;\frac23;-\frac{1}{54\beta^6\nu_\mathrm{min}^6}\right)\nonumber\\
&+\frac{3}{2\beta^4\nu_\mathrm{min}^4}\Gamma\left(\frac43\right)\,_1F_1\left(\frac76;\frac43;-\frac{1}{54\beta^6\nu_\mathrm{min}^6}\right)\nonumber\\
&-\frac{1}{2\beta^6\nu_\mathrm{min}^6}\,_2F_2\left(1,\frac32;\frac43,\frac53;-\frac{1}{54\beta^6\nu_\mathrm{min}^6}\right)\nonumber\\
&+\frac{5}{48\beta^8\nu_\mathrm{min}^8}\Gamma\left(\frac53\right)\,_1F_1\left(\frac{11}{6};\frac53;-\frac{1}{54\beta^6\nu_\mathrm{min}^6}\right)\nonumber\\
&-\frac{7}{288\beta^{10}\nu_\mathrm{min}^{10}}\Gamma\left(\frac43\right)\,_1F_1\left(\frac{13}{6};\frac73;-\frac{1}{54\beta^6\nu_\mathrm{min}^6}\right)\nonumber\\
&\left.\left.+\frac{3}{640\beta^{12}\nu_\mathrm{min}^{12}}\,_2F_2\left(2,\frac52;\frac73,\frac83;-\frac{1}{54\beta^6\nu_\mathrm{min}^6}\right)\right]\right\}^{1/2},
\end{align}
where the local minimum $\nu_\mathrm{min}$ is determined by $U'(\nu)=0$ as
\begin{align}
\frac{1}{\nu_\mathrm{min}^4}-1&=\sqrt{\frac{2}{\pi}}\frac{Na_0^G}{3a_0\nu_\mathrm{min}^3}\left[1\bigg.\right.\nonumber\\
&-\frac{1}{\beta^2\nu_\mathrm{min}^2}\Gamma\left(\frac23\right)\,_1F_1\left(\frac56;\frac23;-\frac{1}{54\beta^6\nu_\mathrm{min}^6}\right)\nonumber\\
&-\frac{1}{9\beta^4\nu_\mathrm{min}^4}\Gamma\left(-\frac23\right)\,_1F_1\left(\frac76;\frac43;-\frac{1}{54\beta^6\nu_\mathrm{min}^6}\right)\nonumber\\
&\left.-\frac{1}{8\beta^6\nu_\mathrm{min}^6}\,_2F_2\left(1,\frac32;\frac43,\frac53;-\frac{1}{54\beta^6\nu_\mathrm{min}^6}\right)\right].
\end{align}

\section{Experimental perspective}\label{sec:exp}

To maximize the effect from the gravitational potential on the effective potential \eqref{eq:UeffVG}, $N$ should be sufficiently large to balance the suppression factor $a_0^G/a_0$ (chosen as $3\times10^{-22}$ specifically in this section). Therefore, we will take $N$ as $10^9$, just one order above the experimentally achievable $N\sim10^6-10^8$ currently, which might be feasible in the far future in space, although it is regarded here just as a preliminary theoretical perspective. We will also take $N$ up to $10^{12}$, although it is not feasible in practice. The smaller number of $N$ could be compensated by a larger ratio of $a_0^G/a_0$ with the larger atomic mass  $m$ and the smaller trapping frequency $\omega_0$, although the two conditions are usually difficult to be satisfied simultaneously in the experiment. 

\begin{figure*}
\centering
\includegraphics[width=0.48\textwidth]{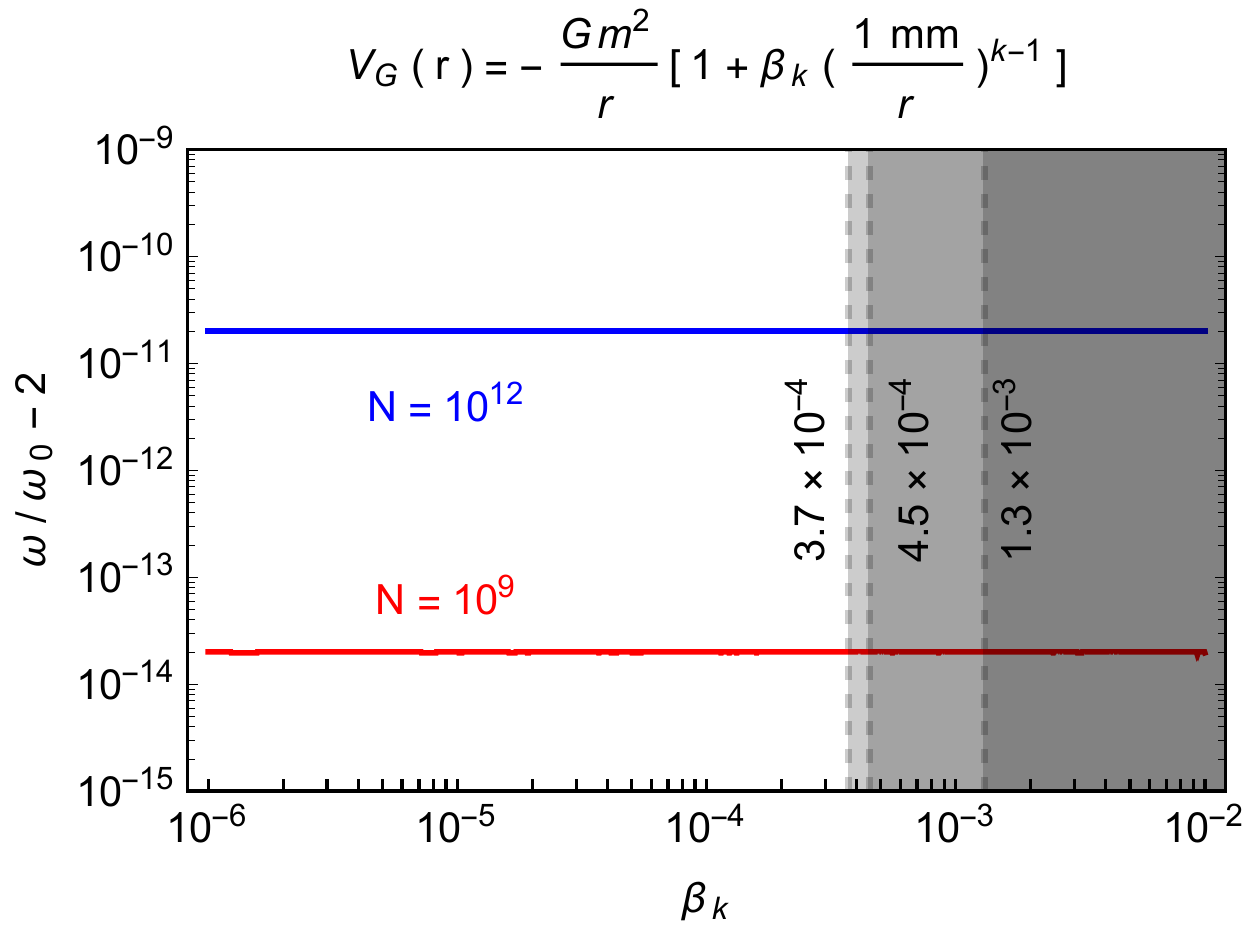}
\includegraphics[width=0.48\textwidth]{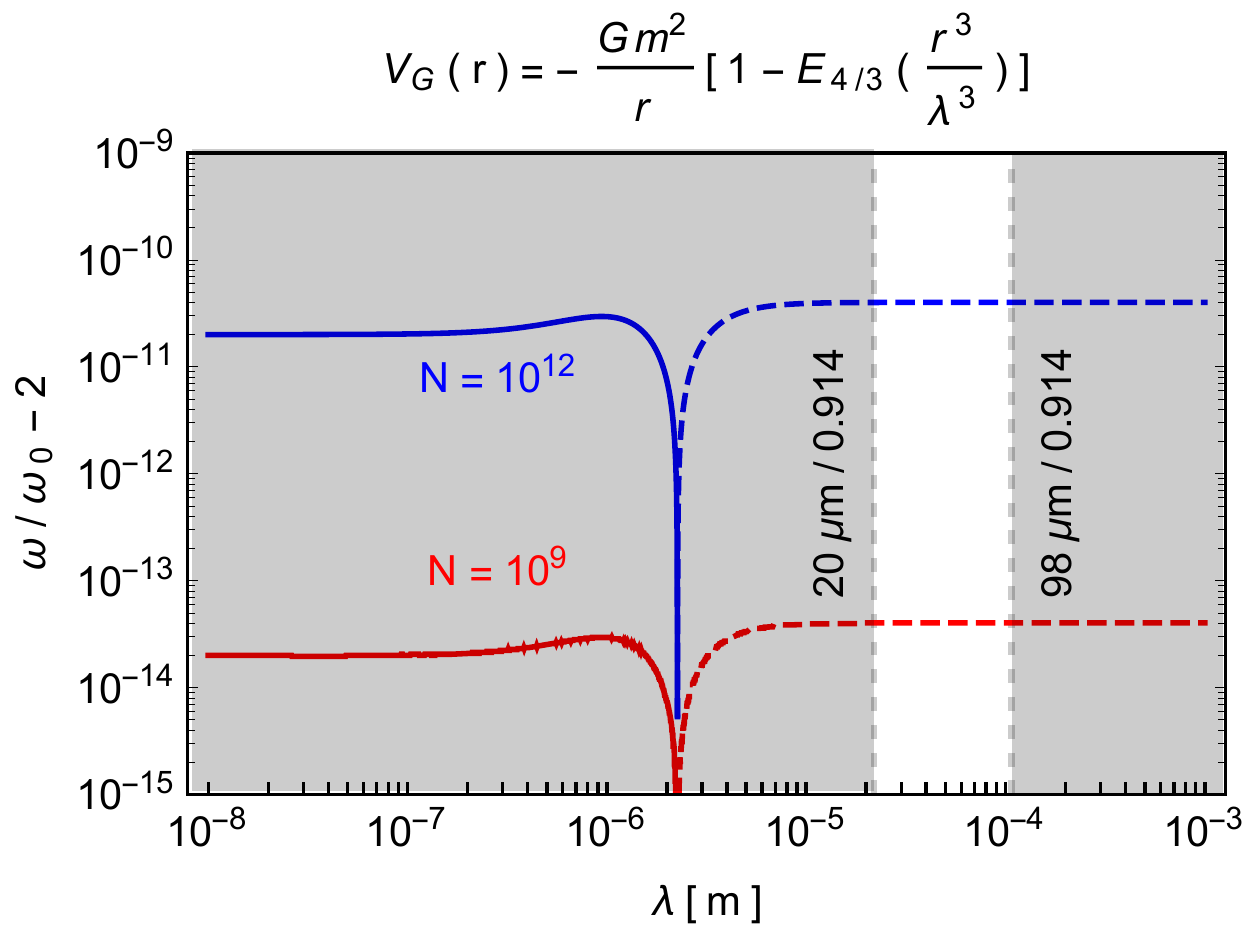}\\
\caption{The frequency deviation in the shape oscillation with respect to the gravity-free case for the power-law potential (left panel) and fat-graviton potential (right panel) with particles number $N=10^9$ (red solid lines) and $N=10^{12}$ (blue solid lines) beyond the current achievable values $N\sim10^6-10^8$. The shaded regions are excluded by current experiments (left panel and $\lambda>98\,\mu\mathrm{m}/0.914$ in the right panel) and naturalness argument ($\lambda<20\,\mu\mathrm{m}/0.914$ in the right panel). The dashed lines are for negative values of $\omega/\omega_0-2$.}\label{fig:Ukf}
\end{figure*}

\begin{figure*}
\centering
\includegraphics[width=0.47\textwidth]{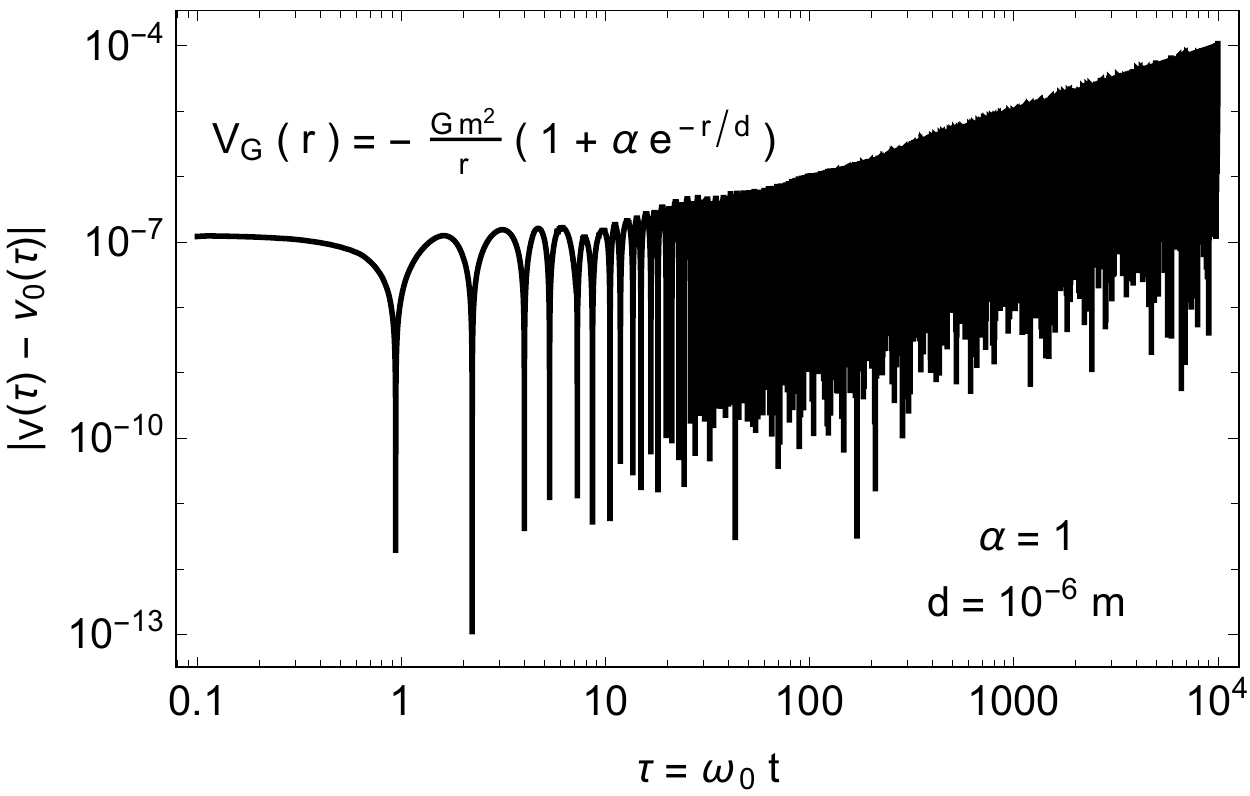}
\includegraphics[width=0.5\textwidth]{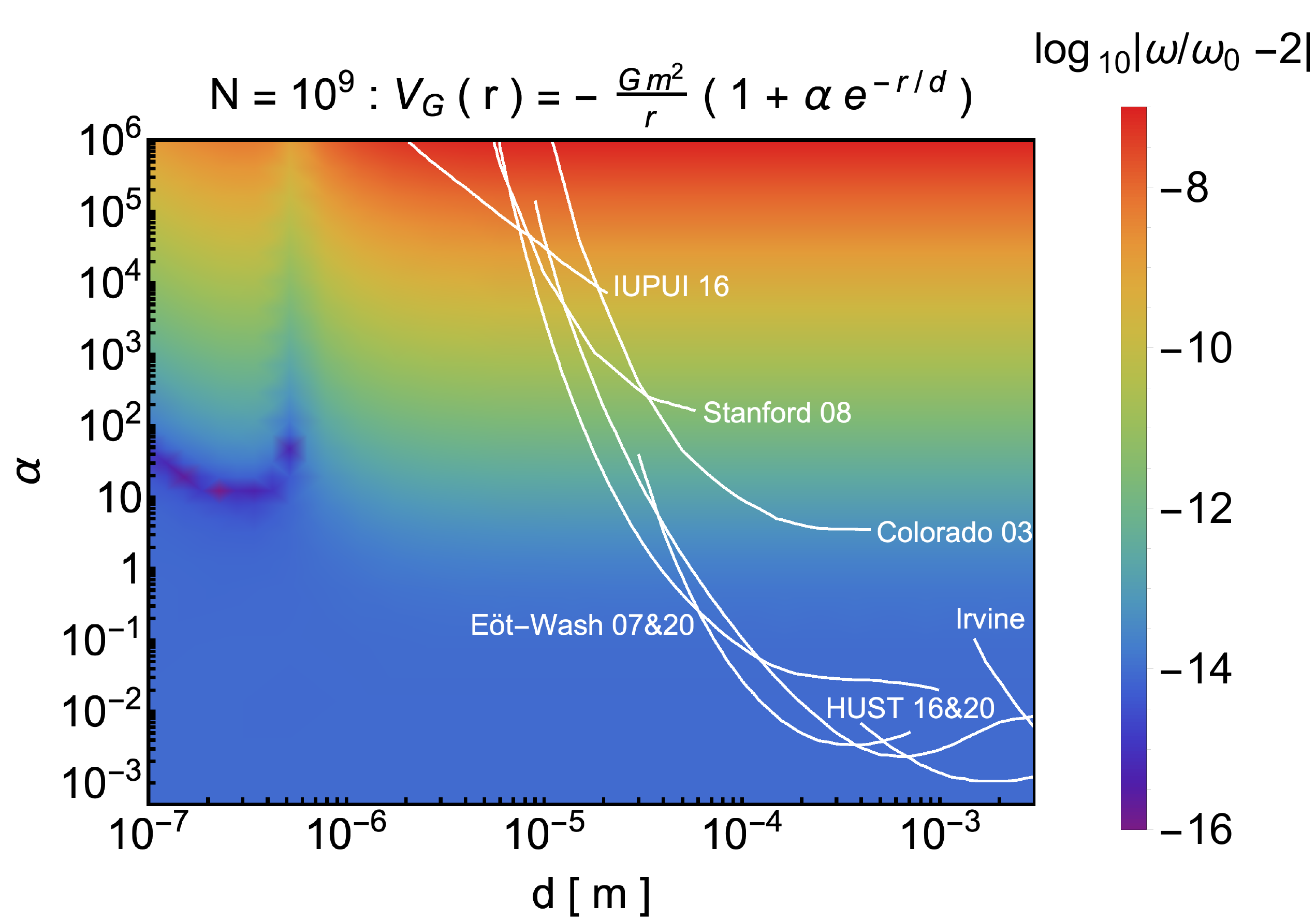}\\
\includegraphics[width=0.47\textwidth]{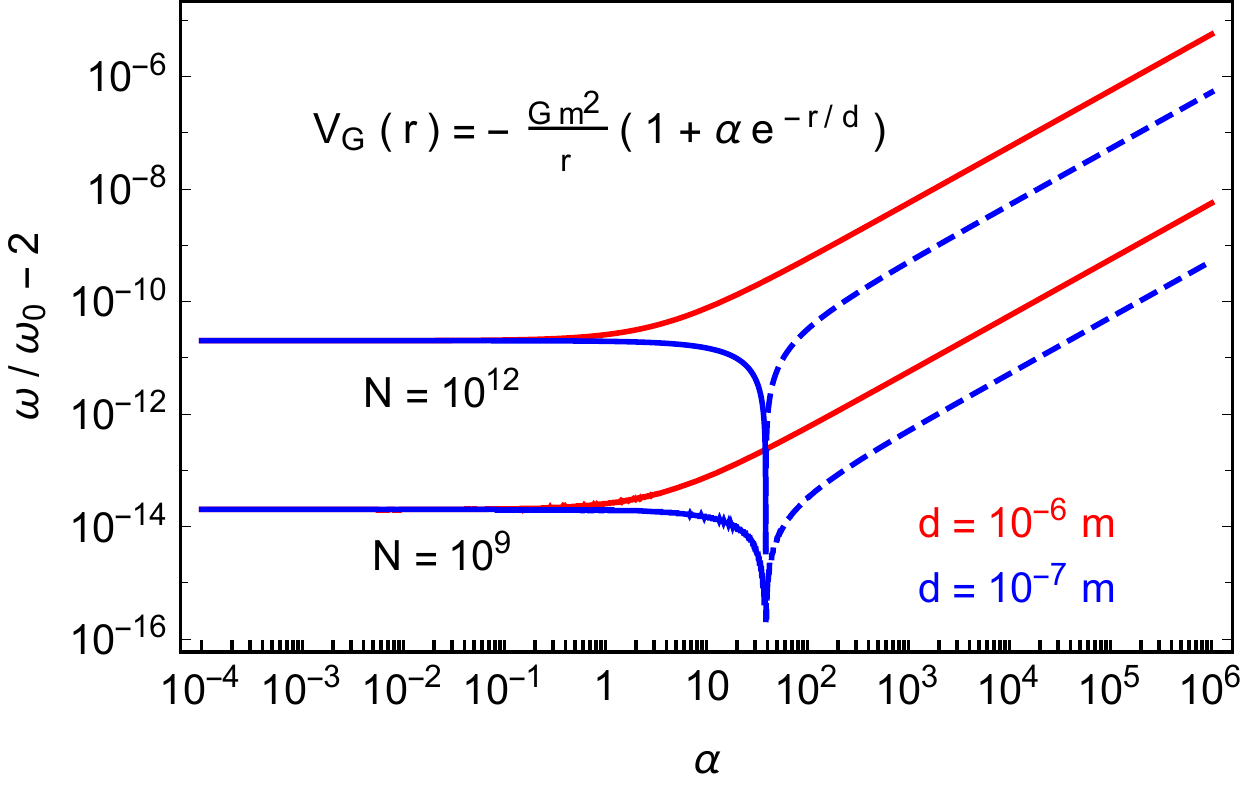}
\includegraphics[width=0.5\textwidth]{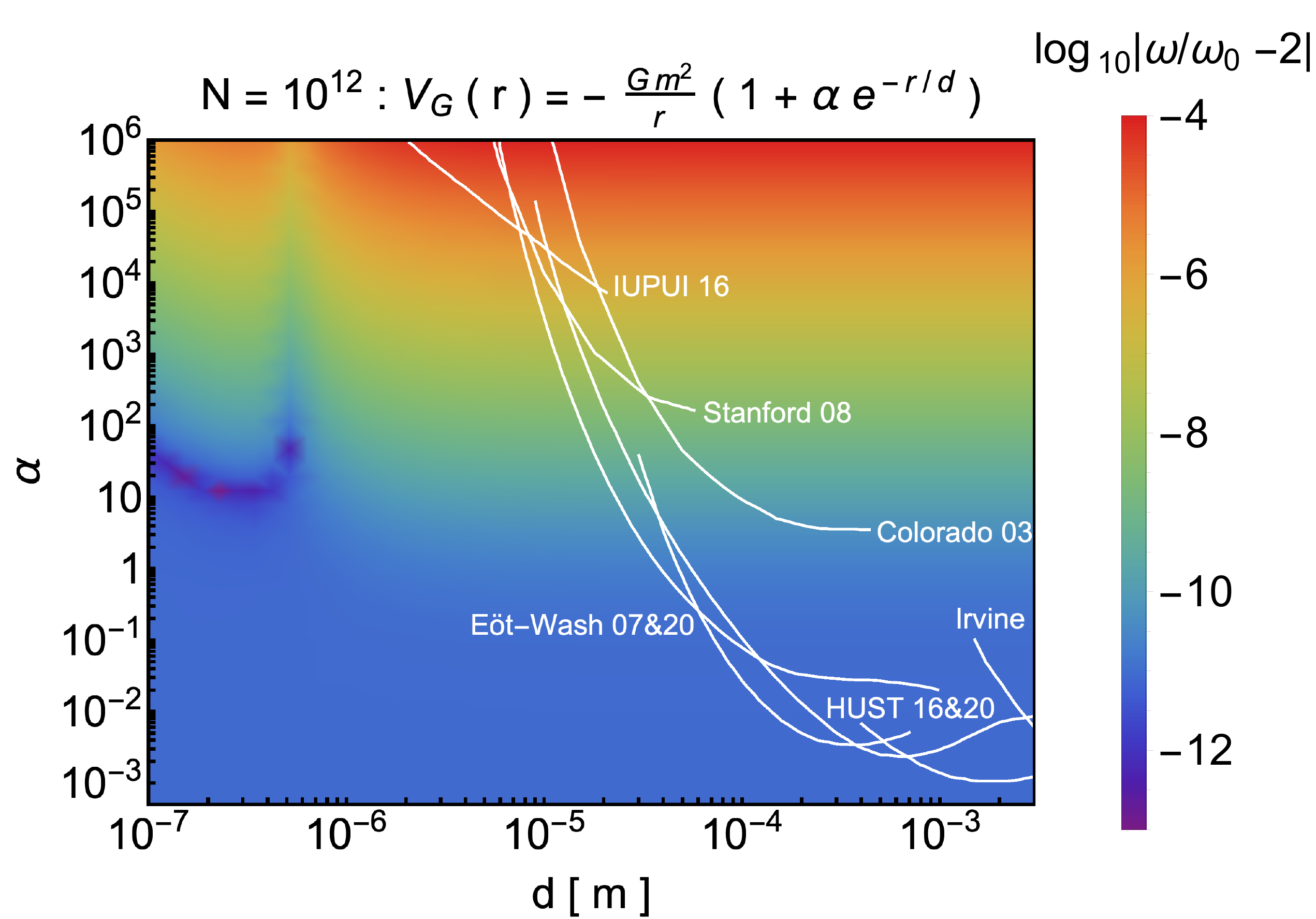}\\
\includegraphics[width=0.45\textwidth]{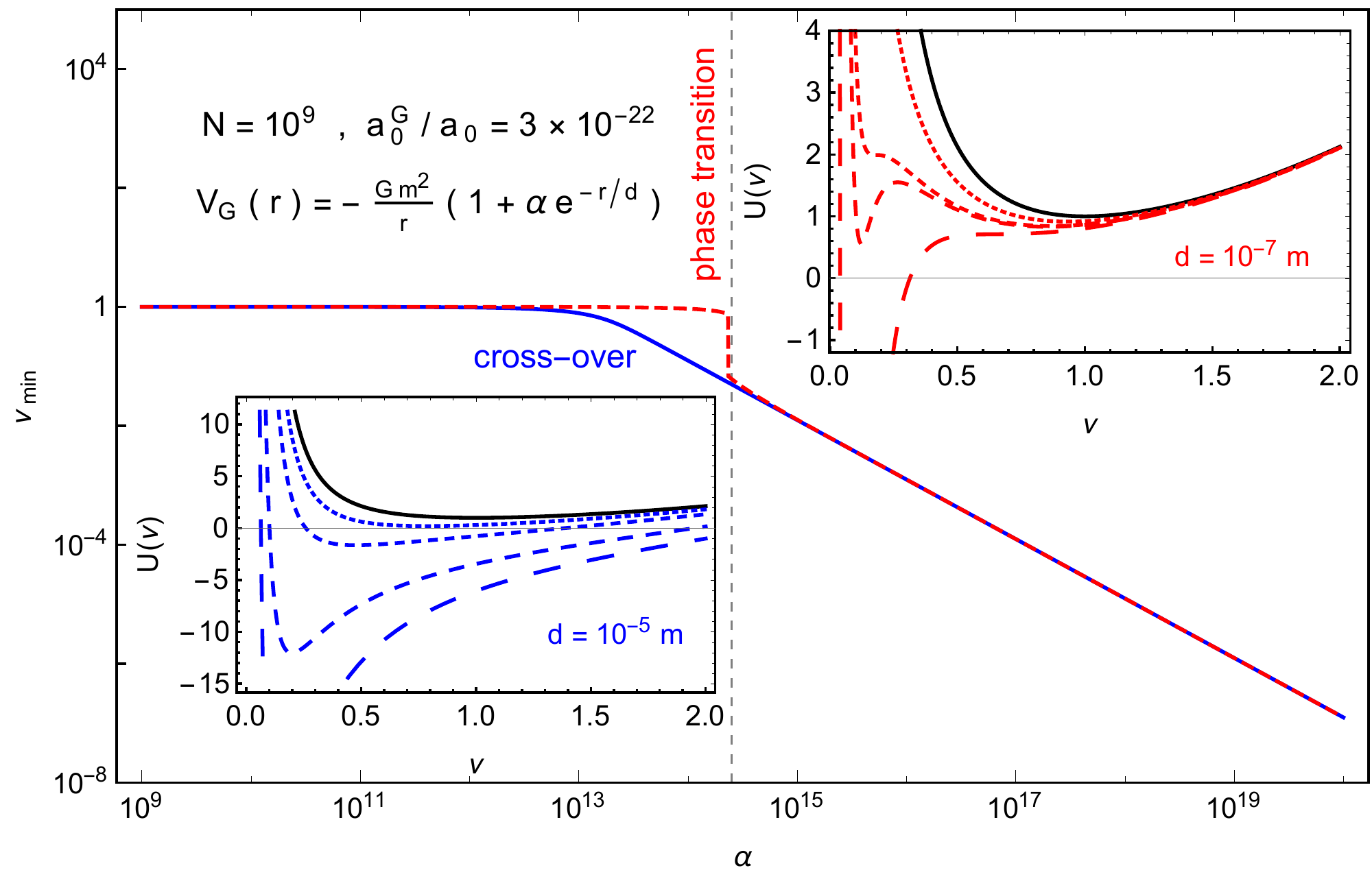}
\includegraphics[width=0.52\textwidth]{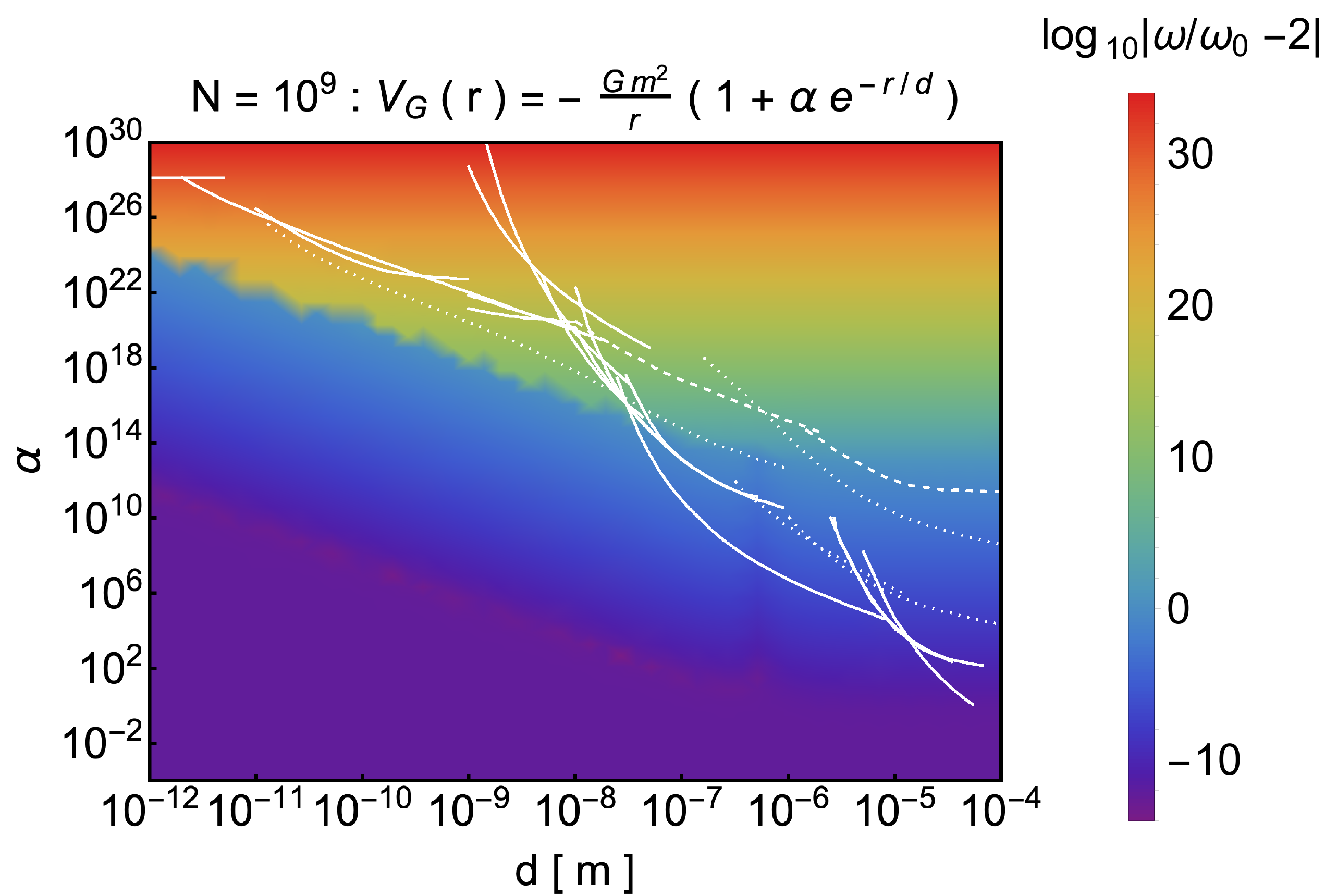}\\
\caption{Experimental perspective for the modified Newtonian gravitational potential with Yukawa interaction. In the first panel, with small deviation in the oscillation frequency with respect to the gravity-free case, the time evolution of the BEC radius deviates with power-law in time.  The absolute value of oscillation frequency deviations with respect to the parameter space of Yukawa interaction are shown in the second and forth panels for different values of condensed particle number with the white curves indicated by the current experimental exclusion limits. There is a small parameter region in the upper left corner with flipped sign for the frequency deviation when the Yuakwa interaction range comparable or smaller than the characteristic length scale of the BEC state as also shown in the third panel. The dashed lines are for negative values of $\omega/\omega_0-2$. In the fifth panel, the potential minimum $\nu_\mathrm{min}$ is depicted with respect to $\alpha$ for different interaction range $d$, which experiences cross-over transition for $d>a_0$ (blue curves) but first-order phase transition for $d<a_0$ (red curves). In the last panel, the absolute deviation of oscillation frequency is presented with respect to a largely extended parameter space supported by $d$ down to $10^{-12}$~m and $\alpha$ up to $10^{30}$. The white curves are borrowed from \cite{ANTONIADIS2011755,Mostepanenko:2020lqe} for comparison with experimental constraints. }\label{fig:UY}
\end{figure*}

\subsection{Power-law potential}

For the power-law potential \eqref{eq:power}, we will take $k=2$ for illustration. The deviation of the oscillation frequency \eqref{eq:omegapower} with respect to the gravity-free oscillation frequency $2\omega_0$ is shown in the left panel of Fig. \ref{fig:Ukf} for the  particle numbers $N=10^9$ (red line) and $N=10^{12}$ (blue line), where the shaded regions $\beta_2>1.3\times10^{-3}$ \cite{Hoskins:1985tn}, $\beta_2>4.5\times10^{-4}$ \cite{Adelberger:2006dh}, and $\beta_2>3.7\times10^{-4}$ \cite{Tan:2020vpf} are excluded by the existing  experiments. Unfortunately, despite of the smallness of frequency deviation, it is also insensitive to the size of coefficient $\beta_k$. Therefore, the ultra-cold BEC experiment in space cannot constrain the modified gravity with power-law gravitational potential.

\subsection{Yukawa interaction}

The experimental perspective for the gravitational potential with Yukawa interaction \eqref{eq:Yukawa} are presented in Fig. \ref{fig:UY}. The deviations of the oscillation frequency \eqref{eq:omegaYukawa} with respect to the gravity-free oscillation frequency $2\omega_0$ are shown for the particle numbers $N=10^9$ (the second panel) and $N=10^{12}$ (the fourth panel) with respect to the current exclusion limits \cite{Hoyle:2004cw,Kapner:2006si,Hoskins:1985tn,Decca:2005qz,Chen:2014oda,Chiaverini:2002cb,Geraci:2008hb,Long:2002wn,Tu:2007zz,Yang:2012zzb,Tan:2016vwu,Tan:2020vpf,Lee:2020zjt}. The frequency deviations are insensitive to both $d$ and $\alpha$ when $\alpha\lesssim\mathcal{O}(10)$, which are as small as $10^{-14}$ (the second panel) and $10^{-11}$ (the fourth panel), respectively. For $d\gtrsim10^{-6}\,\mathrm{m}$, the frequency deviation grows with power-law in the range of $\alpha\gtrsim\mathcal{O}(10)$. For $d<10^{-6}\,\mathrm{m}$, the frequency deviation flips a sign around $\alpha\sim\mathcal{O}(10)$, after which the absolute value of frequency deviation still grows with power-law in $\alpha$. This could also be seen in the third panel for some illustrative values of $N$ and $d$, where the dashed lines denote for the negative value of the frequency deviation $\omega/\omega_0-2$. Note that the regime of flipped sign appears when $d$ is smaller than the characteristic length scale $a_0\simeq10^{-6}\,\mathrm{m}$ of the BEC state. Except for the region of flipped sign, the frequency deviation would be insensitive to the interaction range $d$ for given $\alpha$, which could greatly push current exclusion limits.

To detect such a small frequency deviation, for example in the regime of $\alpha\lesssim\mathcal{O}(10)$ with $\omega/\omega_0-2\simeq10^{-14}$ for $N=10^9$ and $\omega/\omega_0-2\simeq10^{-11}$ for $N=10^{12}$ , we need the resolution of frequency in the measurement as
\begin{align}
N=10^{9} &: \Delta\omega\simeq10^{-14}\omega\simeq(1-100)\,\mathrm{pHz},\\
N=10^{12} &: \Delta\omega\simeq10^{-11}\omega\simeq(1-100)\,\mathrm{nHz},
\end{align}
where the trapping frequency is usually in the range of $50-10,000$ Hz. The above requirement for the resolution of frequency could be either relaxed in the regime where the absolute value of frequency deviation $|\omega/\omega_0-2|$ grows with the power-law in $\alpha$, or enhanced in the regime where the frequency deviation $\omega/\omega_0-2$ flips a sign. 

However, the actual requirement for the resolution of frequency measurement could be further relaxed universally in the whole parameter space of $d$ and $\alpha$ by virtue of the long lifetime of ultra-cold BEC in space. In fact, the difference between the measured time-evolution $\nu(\tau)$ of oscillating BEC and the gravity-free solution $\nu_0(\tau)$ grows with power-law in time $\tau$ as seen in the first panel of Fig. \ref{fig:UY}. Furthermore, for trapping frequency in the range of $50-10,000$ Hz, the measurement of BEC oscillations up to $\tau\sim10^4$ corresponds to a condensate lifetime of $1-100$ seconds that is exactly comparable to the free-fall time achieved with the help of microgravity environment. Therefore, even if the current resolution of frequency-measurement is not high enough to determine the oscillation frequency within a single oscillation period, the long-time measurement of oscillating BEC in space allows us to calibrate the measured frequency from previous oscillation period to higher precision over the next numerous oscillation periods. In our specific case, the measurement of oscillations up to $\tau\sim10^4$ corresponds to  $n\sim10^3$ periods, which could further relax the resolution of frequency measurement down to
\begin{align}
N=10^9 &: \Delta\omega_{nT}\sim n\Delta\omega\simeq(1-100)\,\mathrm{nHz},\\
N=10^{12} &: \Delta\omega_{nT}\sim n\Delta\omega\simeq(1-100)\,\mu\mathrm{Hz}.
\end{align}

As matter of a preliminary investigation, we also study the Yukawa interaction in a largely extended parameter space supported by the interaction range $d$ down to $10^{-12}$~m and strength $\alpha$ up to $10^{30}$. For the cases of $d>a_0$ and $d<a_0$, the potential minimum $\nu_\mathrm{min}$ experiences a cross-over and first-order phase transitions with increasing $\alpha$ as shown with blue and red curves in the fifth panel of Fig. \ref{fig:UY}, respectively. The absolute deviation of oscillation frequency in this extended parameter space is depicted in the last panel of Fig. \ref{fig:UY}, which is relatively small for the blue and purple regions so that the impact on the relative error of the oscillation frequency from the Newtonian gravitational potential is negligible compared to the measured deviation of oscillation frequency according to \eqref{eq:dwbyw}. Therefore, this parameter regions could be promising to be ruled out by future ultra-cold atom experiments. The rest of the parameter regions, although largely affected by the Newtonian potential, were already ruled out mostly by the current constraints reviewed in  \cite{ANTONIADIS2011755,Mostepanenko:2020lqe} and references therein. Furthermore, the idealistic condition we assume for neglecting the second term in \eqref{eq:UeffVG} could be relaxed for large $\alpha$ in the extended parameter space, and the further investigation for the full competitions of all three terms in \eqref{eq:UeffVG} will be reserved for future work.

\subsection{Fat graviton}

For the fat-graviton potential \eqref{eq:fat}, the deviation of oscillation frequency \eqref{eq:omegafat} with respect to the gravity-free oscillation frequency $2\omega_0$ is shown in the right panel of Fig. \ref{fig:Ukf} for the particle numbers $N=10^9$ (red line) and $N=10^{12}$ (blue line), where the dashed lines denote for the negative value of the frequency deviation $\omega/\omega_0-2$, and the shaded region $\lambda>98\,\mu\mathrm{m}/0.914$ is excluded by the experiments \cite{Kapner:2006si,Adelberger:2006dh}. The other shaded region $\lambda<20\,\mu\mathrm{m}/0.914$ is argued by naturalness in \cite{Sundrum:2003jq}. Similar to case with power-law potential, the frequency deviation is also insensitive to $\lambda$ regardless of the flipped sign around $\lambda\simeq2\times10^{-6}\,\mathrm{m}$, which is also roughly the characteristic length scale $a_0\simeq10^{-6}\,\mathrm{m}$ of BEC state. Therefore, the ultra-cold BEC experiment in space also cannot constrain the modified gravity with fat-graviton potential in the regime of interest $20\,\mu\mathrm{m}/0.914<\lambda<98\,\mu\mathrm{m}/0.914$.

\section{Conclusion and discussions}\label{sec:con}

Contrary to the native expectation that gravity cannot be further constrained at small scales since it is weakest among all other fundamental/quintessential interactions, the ultra-cold BEC experiments in space with microgravity environment could explore the parameter space in the modified Newtonian potentials because this experiment could significantly extend the observation time so as to precisely measure the microscopic properties of gravity via its macroscopic manifestation on the BEC state with shape oscillation. Newtonian gravity and  several modified gravity theories are examined in detail. The results with Yukawa interaction are optimistic in probing the parameter space greatly beyond the current experimental exclusion limits. However, the results with the Newtonian gravity, power-law potential and fat-graviton potential are quite pessimistic. Several comments are given below:

Firstly, our proposal of measurement on Newtonian gravitational constant is new to our knowledge in the sense of the use of cold-atom BEC with shape oscillation, although it is experimentally more challenging than the current use of atom interferometer.

Secondly, the long lifetime of BEC state in space buys us more time to measure and calibrate the oscillation frequency over numerous periods so that the actual resolution of frequency measurement could be relaxed compared to the frequency resolution in a single period.

Thirdly, our estimation for the frequency resolution is quite conservative, which could be further relaxed for $\alpha\gtrsim\mathcal{O}(10)$ in the regime of $d\sim10^{-6}-10^{-5}\,\mathrm{m}$ where frequency deviation grows with power-law in $\alpha$. This greatly improves the current exclusion limits.

\begin{acknowledgments}
We would like to thank Run-Qiu Yang for useful correspondences.
Rong-Gen Cai is supported by the National Natural Science Foundation of China Grants No. 11947302, No. 11991052, No. 11690022, No. 11821505 and No. 11851302, and by the Strategic Priority Research Program of CAS Grant No. XDB23030100, and by the Key Research Program of Frontier Sciences of CAS. Su Yi is supported by the National Key Research and Development Program of China (Grant No. 2017YFA0304501). Jiang-Hao Yu is supported by the National Science Foundation of China under Grants No. 11875003 and No. 11947302. Shao-Jiang Wang was supported by the postdoctoral scholarship of Tufts University from NSF when part of the work was done at Tufts University. 
\end{acknowledgments}


\bibliographystyle{utphys}
\bibliography{ref}

\providecommand{\href}[2]{#2}\begingroup\raggedright\begin{thebibliography}{10}

\bibitem{ArkaniHamed:2006dz}
N.~Arkani-Hamed, L.~Motl, A.~Nicolis, and C.~Vafa, ``{The String landscape,
  black holes and gravity as the weakest force},''
  \href{http://dx.doi.org/10.1088/1126-6708/2007/06/060}{{\em JHEP} {\bfseries
  06} (2007) 060},
\href{http://arxiv.org/abs/hep-th/0601001}{{\ttfamily arXiv:hep-th/0601001
  [hep-th]}}.

\bibitem{Palti:2017elp}
E.~Palti, ``{The Weak Gravity Conjecture and Scalar Fields},''
  \href{http://dx.doi.org/10.1007/JHEP08(2017)034}{{\em JHEP} {\bfseries 08}
  (2017) 034},
\href{http://arxiv.org/abs/1705.04328}{{\ttfamily arXiv:1705.04328 [hep-th]}}.

\bibitem{Gonzalo:2019gjp}
E.~Gonzalo and L.~E. Ibáñez, ``{A Strong Scalar Weak Gravity Conjecture and
  Some Implications},'' \href{http://dx.doi.org/10.1007/JHEP08(2019)118}{{\em
  JHEP} {\bfseries 08} (2019) 118},
\href{http://arxiv.org/abs/1903.08878}{{\ttfamily arXiv:1903.08878 [hep-th]}}.

\bibitem{Cai:2019dzj}
R.-G. Cai and S.-J. Wang, ``{A refined trans-Planckian censorship
  conjecture},'' \href{http://arxiv.org/abs/1912.00607}{{\ttfamily
  arXiv:1912.00607 [hep-th]}}.

\bibitem{Cheung:2018cwt}
C.~Cheung, J.~Liu, and G.~N. Remmen, ``{Proof of the Weak Gravity Conjecture
  from Black Hole Entropy},''
  \href{http://dx.doi.org/10.1007/JHEP10(2018)004}{{\em JHEP} {\bfseries 10}
  (2018) 004},
\href{http://arxiv.org/abs/1801.08546}{{\ttfamily arXiv:1801.08546 [hep-th]}}.

\bibitem{Adelberger:2009zz}
E.~G. Adelberger, J.~H. Gundlach, B.~R. Heckel, S.~Hoedl, and S.~Schlamminger,
  ``{Torsion balance experiments: A low-energy frontier of particle physics},''
\href{http://dx.doi.org/10.1016/j.ppnp.2008.08.002}{{\em Prog. Part. Nucl.
  Phys.} {\bfseries 62} (2009) 102--134}.

\bibitem{Uzan:2010ri}
J.-P. Uzan, ``{Tests of General Relativity on Astrophysical Scales},''
  \href{http://dx.doi.org/10.1007/s10714-010-1047-8}{{\em Gen. Rel. Grav.}
  {\bfseries 42} (2010) 2219--2246},
\href{http://arxiv.org/abs/0908.2243}{{\ttfamily arXiv:0908.2243
  [astro-ph.CO]}}.

\bibitem{Jain:2010ka}
B.~Jain and J.~Khoury, ``{Cosmological Tests of Gravity},''
  \href{http://dx.doi.org/10.1016/j.aop.2010.04.002}{{\em Annals Phys.}
  {\bfseries 325} (2010) 1479--1516},
\href{http://arxiv.org/abs/1004.3294}{{\ttfamily arXiv:1004.3294
  [astro-ph.CO]}}.

\bibitem{Baker:2014zba}
T.~Baker, D.~Psaltis, and C.~Skordis, ``{Linking Tests of Gravity On All
  Scales: from the Strong-Field Regime to Cosmology},''
  \href{http://dx.doi.org/10.1088/0004-637X/802/1/63}{{\em Astrophys. J.}
  {\bfseries 802} (2015) 63},
\href{http://arxiv.org/abs/1412.3455}{{\ttfamily arXiv:1412.3455
  [astro-ph.CO]}}.

\bibitem{Berti:2015itd}
E.~Berti {\em et~al.}, ``{Testing General Relativity with Present and Future
  Astrophysical Observations},''
  \href{http://dx.doi.org/10.1088/0264-9381/32/24/243001}{{\em Class. Quant.
  Grav.} {\bfseries 32} (2015) 243001},
\href{http://arxiv.org/abs/1501.07274}{{\ttfamily arXiv:1501.07274 [gr-qc]}}.

\bibitem{Sakstein:2015oqa}
J.~Sakstein, \href{http://dx.doi.org/10.17863/CAM.16133}{{\em {Astrophysical
  Tests of Modified Gravity}}}.
\newblock PhD thesis, Cambridge U., DAMTP, 2014.
\newblock \href{http://arxiv.org/abs/1502.04503}{{\ttfamily arXiv:1502.04503
  [astro-ph.CO]}}.
\newblock
\url{https://www.repository.cam.ac.uk/handle/1810/246265}.
\newblock

\bibitem{Koyama:2015vza}
K.~Koyama, ``{Cosmological Tests of Modified Gravity},''
  \href{http://dx.doi.org/10.1088/0034-4885/79/4/046902}{{\em Rept. Prog.
  Phys.} {\bfseries 79} no.~4, (2016) 046902},
\href{http://arxiv.org/abs/1504.04623}{{\ttfamily arXiv:1504.04623
  [astro-ph.CO]}}.

\bibitem{Ishak:2018his}
M.~Ishak, ``{Testing General Relativity in Cosmology},''
  \href{http://dx.doi.org/10.1007/s41114-018-0017-4}{{\em Living Rev. Rel.}
  {\bfseries 22} no.~1, (2019) 1},
\href{http://arxiv.org/abs/1806.10122}{{\ttfamily arXiv:1806.10122
  [astro-ph.CO]}}.

\bibitem{Ferreira:2019xrr}
P.~G. Ferreira, ``{Cosmological Tests of Gravity},''
  \href{http://dx.doi.org/10.1146/annurev-astro-091918-104423}{{\em Ann. Rev.
  Astron. Astrophys.} {\bfseries 57} (2019) 335--374},
\href{http://arxiv.org/abs/1902.10503}{{\ttfamily arXiv:1902.10503
  [astro-ph.CO]}}.

\bibitem{Baker:2019gxo}
T.~Baker {\em et~al.}, ``{The Novel Probes Project -- Tests of Gravity on
  Astrophysical Scales},''
\href{http://arxiv.org/abs/1908.03430}{{\ttfamily arXiv:1908.03430
  [astro-ph.CO]}}.

\bibitem{vanZoest1540}
T.~van Zoest, N.~Gaaloul, Y.~Singh, H.~Ahlers, W.~Herr, S.~T. Seidel,
  W.~Ertmer, E.~Rasel, M.~Eckart, E.~Kajari, S.~Arnold, G.~Nandi, W.~P.
  Schleich, R.~Walser, A.~Vogel, K.~Sengstock, K.~Bongs, W.~Lewoczko-Adamczyk,
  M.~Schiemangk, T.~Schuldt, A.~Peters, T.~K{\"o}nemann, H.~M{\"u}ntinga,
  C.~L{\"a}mmerzahl, H.~Dittus, T.~Steinmetz, T.~W. H{\"a}nsch, and J.~Reichel,
  ``Bose-einstein condensation in microgravity,''
  \href{http://dx.doi.org/10.1126/science.1189164}{{\em Science} {\bfseries
  328} no.~5985, (2010) 1540--1543}.
  \url{https://science.sciencemag.org/content/328/5985/1540}.

\bibitem{Geiger2011}
R.~Geiger, V.~Ménoret, G.~Stern, N.~Zahzam, P.~Cheinet, B.~Battelier,
  A.~Villing, F.~Moron, M.~Lours, Y.~Bidel, A.~Bresson, A.~Landragin, and
  P.~Bouyer, ``Detecting inertial effects with airborne matter-wave
  interferometry,'' \href{http://dx.doi.org/10.1038/ncomms1479}{{\em Nat.
  Commun.} {\bfseries 2} (2011) 474}. \url{https://doi.org/10.1038/ncomms1479}.

\bibitem{Becker2018}
D.~Becker, M.~D. Lachmann, S.~T. Seidel, H.~Ahlers, A.~N. Dinkelaker,
  J.~Grosse, O.~Hellmig, H.~Müntinga, V.~Schkolnik, T.~Wendrich,
  A.~Wenzlawski, B.~Weps, R.~Corgier, T.~Franz, N.~Gaaloul, W.~Herr,
  D.~Lüdtke, M.~Popp, S.~Amri, H.~Duncker, M.~Erbe, A.~Kohfeldt,
  A.~Kubelka-Lange, C.~Braxmaier, E.~Charron, W.~Ertmer, M.~Krutzik,
  C.~Lämmerzahl, A.~Peters, W.~P. Schleich, K.~Sengstock, R.~Walser, A.~Wicht,
  P.~Windpassinger, and E.~M. Rasel, ``Space-borne bose–einstein condensation
  for precision interferometry,''
  \href{http://dx.doi.org//10.1038/s41586-018-0605-1}{{\em Nature} {\bfseries
  562} (2018) 391–395}. \url{https://doi.org/10.1038/s41586-018-0605-1}.

\bibitem{Tino:2020dsl}
G.~M. Tino, ``{Testing gravity with cold atom interferometry: Results and
  prospects},'' \href{http://dx.doi.org/10.1088/2058-9565/abd83e}{{\em Quantum
  Sci. Technol.} {\bfseries 6} no.~2, (2021) 024014},
  \href{http://arxiv.org/abs/2009.01484}{{\ttfamily arXiv:2009.01484 [gr-qc]}}.

\bibitem{Bertoldi:2019tck}
{\bfseries AEDGE} Collaboration, Y.~A. El-Neaj {\em et~al.}, ``{AEDGE: Atomic
  Experiment for Dark Matter and Gravity Exploration in Space},''
  \href{http://dx.doi.org/10.1140/epjqt/s40507-020-0080-0}{{\em EPJ Quant.
  Technol.} {\bfseries 7} (2020) 6},
\href{http://arxiv.org/abs/1908.00802}{{\ttfamily arXiv:1908.00802 [gr-qc]}}.

\bibitem{Adelberger:2006dh}
E.~G. Adelberger, B.~R. Heckel, S.~A. Hoedl, C.~D. Hoyle, D.~J. Kapner, and
  A.~Upadhye, ``{Particle Physics Implications of a Recent Test of the
  Gravitational Inverse Sqaure Law},''
  \href{http://dx.doi.org/10.1103/PhysRevLett.98.131104}{{\em Phys. Rev. Lett.}
  {\bfseries 98} (2007) 131104},
\href{http://arxiv.org/abs/hep-ph/0611223}{{\ttfamily arXiv:hep-ph/0611223
  [hep-ph]}}.

\bibitem{su:93}
J.~Sucher and G.~Feinberg, {\em Long-Range Casimir Forces}.
\newblock Plenum, New York, f.s. levin and d.a. micha~ed., 1993.

\bibitem{Ferrer:1998rw}
F.~Ferrer and M.~Nowakowski, ``{Higgs and Goldstone bosons mediated long range
  forces},'' \href{http://dx.doi.org/10.1103/PhysRevD.59.075009}{{\em Phys.
  Rev. D} {\bfseries 59} (1999) 075009},
  \href{http://arxiv.org/abs/hep-ph/9810550}{{\ttfamily arXiv:hep-ph/9810550}}.

\bibitem{Deshpande:2007mf}
N.~Deshpande, S.~D. Hsu, and J.~Jiang, ``{Long range forces and limits on
  unparticle interactions},''
  \href{http://dx.doi.org/10.1016/j.physletb.2007.12.018}{{\em Phys. Lett. B}
  {\bfseries 659} (2008) 888--890},
  \href{http://arxiv.org/abs/0708.2735}{{\ttfamily arXiv:0708.2735 [hep-ph]}}.

\bibitem{Sundrum:2003jq}
R.~Sundrum, ``{Fat gravitons, the cosmological constant and submillimeter
  tests},'' \href{http://dx.doi.org/10.1103/PhysRevD.69.044014}{{\em Phys. Rev.
  D} {\bfseries 69} (2004) 044014},
  \href{http://arxiv.org/abs/hep-th/0306106}{{\ttfamily arXiv:hep-th/0306106}}.

\bibitem{PhysRevLett.77.5320}
V.~M. P\'erez-Garc\'{\i}a, H.~Michinel, J.~I. Cirac, M.~Lewenstein, and
  P.~Zoller, ``Low energy excitations of a bose-einstein condensate: A
  time-dependent variational analysis,''
  \href{http://dx.doi.org/10.1103/PhysRevLett.77.5320}{{\em Phys. Rev. Lett.}
  {\bfseries 77} (Dec, 1996) 5320--5323}.
  \url{https://link.aps.org/doi/10.1103/PhysRevLett.77.5320}.

\bibitem{PhysRevA.56.1424}
V.~M. P\'erez-Garc\'{\i}a, H.~Michinel, J.~I. Cirac, M.~Lewenstein, and
  P.~Zoller, ``Dynamics of bose-einstein condensates: Variational solutions of
  the gross-pitaevskii equations,''
  \href{http://dx.doi.org/10.1103/PhysRevA.56.1424}{{\em Phys. Rev. A}
  {\bfseries 56} (Aug, 1997) 1424--1432}.
  \url{https://link.aps.org/doi/10.1103/PhysRevA.56.1424}.

\bibitem{Gupta:2015cta}
P.~Das~Gupta, ``{Gravity, Bose-Einstein Condensates and Gross-Pitaevskii
  Equation},''
\newblock 5, 2015.
\newblock \href{http://arxiv.org/abs/1505.00541}{{\ttfamily arXiv:1505.00541
  [gr-qc]}}.

\bibitem{Cornish:2000zz}
S.~L. Cornish, N.~R. Claussen, J.~L. Roberts, E.~A. Cornell, and C.~E. Wieman,
  ``{Stable Rb-85 Bose-Einstein Condensates with Widely Tunable
  Interactions},'' \href{http://dx.doi.org/10.1103/PhysRevLett.85.1795}{{\em
  Phys. Rev. Lett.} {\bfseries 85} (2000) 1795--1798},
\href{http://arxiv.org/abs/cond-mat/0004290}{{\ttfamily arXiv:cond-mat/0004290
  [cond-mat]}}.

\bibitem{PhysRevA.47.4114}
E.~Tiesinga, B.~J. Verhaar, and H.~T.~C. Stoof, ``Threshold and resonance
  phenomena in ultracold ground-state collisions,''
  \href{http://dx.doi.org/10.1103/PhysRevA.47.4114}{{\em Phys. Rev. A}
  {\bfseries 47} (May, 1993) 4114--4122}.
  \url{https://link.aps.org/doi/10.1103/PhysRevA.47.4114}.

\bibitem{Inouye1998}
S.~Inouye, M.~R. Andrews, J.~Stenger, H.~J. Miesner, D.~M. Stamper-Kurn, and
  W.~Ketterle, ``Observation of feshbach resonances in a bose--einstein
  condensate,'' \href{http://dx.doi.org/10.1038/32354}{{\em Nature} {\bfseries
  392} no.~6672, (1998) 151--154}. \url{https://doi.org/10.1038/32354}.

\bibitem{PhysRevLett.81.69}
P.~Courteille, R.~S. Freeland, D.~J. Heinzen, F.~A. van Abeelen, and B.~J.
  Verhaar, ``Observation of a feshbach resonance in cold atom scattering,''
  \href{http://dx.doi.org/10.1103/PhysRevLett.81.69}{{\em Phys. Rev. Lett.}
  {\bfseries 81} (Jul, 1998) 69--72}.
  \url{https://link.aps.org/doi/10.1103/PhysRevLett.81.69}.

\bibitem{PhysRevLett.81.5109}
J.~L. Roberts, N.~R. Claussen, J.~P. Burke, C.~H. Greene, E.~A. Cornell, and
  C.~E. Wieman, ``Resonant magnetic field control of elastic scattering in cold
  $^{85}rb$,'' \href{http://dx.doi.org/10.1103/PhysRevLett.81.5109}{{\em Phys.
  Rev. Lett.} {\bfseries 81} (Dec, 1998) 5109--5112}.
  \url{https://link.aps.org/doi/10.1103/PhysRevLett.81.5109}.

\bibitem{Rosi:2014kva}
G.~Rosi, F.~Sorrentino, L.~Cacciapuoti, M.~Prevedelli, and G.~M. Tino,
  ``{Precision Measurement of the Newtonian Gravitational Constant Using Cold
  Atoms},'' \href{http://dx.doi.org/10.1038/nature13433}{{\em Nature}
  {\bfseries 510} (2014) 518},
\href{http://arxiv.org/abs/1412.7954}{{\ttfamily arXiv:1412.7954
  [physics.atom-ph]}}.

\bibitem{Fixler:2007is}
J.~B. Fixler, G.~T. Foster, J.~M. McGuirk, and M.~A. Kasevich, ``{Atom
  interferometer measurement of the Newtonian constant of gravity},''
\href{http://dx.doi.org/10.1126/science.1135459}{{\em Science} {\bfseries 315}
  (2007) 74--77}.

\bibitem{ANTONIADIS2011755}
I.~Antoniadis, S.~Baessler, M.~Büchner, V.~Fedorov, S.~Hoedl, A.~Lambrecht,
  V.~Nesvizhevsky, G.~Pignol, K.~Protasov, S.~Reynaud, and Y.~Sobolev,
  ``Short-range fundamental forces,''
  \href{http://dx.doi.org/https://doi.org/10.1016/j.crhy.2011.05.004}{{\em
  Comptes Rendus Physique} {\bfseries 12} no.~8, (2011) 755--778}.
  \url{https://www.sciencedirect.com/science/article/pii/S1631070511001393}.
  Ultra cold neutron quantum states.

\bibitem{Mostepanenko:2020lqe}
V.~M. Mostepanenko and G.~L. Klimchitskaya, ``{The State of the Art in
  Constraining Axion-to-Nucleon Coupling and Non-Newtonian Gravity from
  Laboratory Experiments},''
  \href{http://dx.doi.org/10.3390/universe6090147}{{\em Universe} {\bfseries 6}
  no.~9, (2020) 147}, \href{http://arxiv.org/abs/2009.04517}{{\ttfamily
  arXiv:2009.04517 [hep-ph]}}.

\bibitem{Hoskins:1985tn}
J.~K. Hoskins, R.~D. Newman, R.~Spero, and J.~Schultz, ``{Experimental tests of
  the gravitational inverse square law for mass separations from 2-cm to
  105-cm},''
\href{http://dx.doi.org/10.1103/PhysRevD.32.3084}{{\em Phys. Rev.} {\bfseries
  D32} (1985) 3084--3095}.

\bibitem{Tan:2020vpf}
W.-H. Tan {\em et~al.}, ``{Improvement for Testing the Gravitational
  Inverse-Square Law at the Submillimeter Range},''
\href{http://dx.doi.org/10.1103/PhysRevLett.124.051301}{{\em Phys. Rev. Lett.}
  {\bfseries 124} no.~5, (2020) 051301}.

\bibitem{Hoyle:2004cw}
C.~D. Hoyle, D.~J. Kapner, B.~R. Heckel, E.~G. Adelberger, J.~H. Gundlach,
  U.~Schmidt, and H.~E. Swanson, ``{Sub-millimeter tests of the gravitational
  inverse-square law},''
  \href{http://dx.doi.org/10.1103/PhysRevD.70.042004}{{\em Phys. Rev.}
  {\bfseries D70} (2004) 042004},
\href{http://arxiv.org/abs/hep-ph/0405262}{{\ttfamily arXiv:hep-ph/0405262
  [hep-ph]}}.

\bibitem{Kapner:2006si}
D.~J. Kapner, T.~S. Cook, E.~G. Adelberger, J.~H. Gundlach, B.~R. Heckel, C.~D.
  Hoyle, and H.~E. Swanson, ``{Tests of the gravitational inverse-square law
  below the dark-energy length scale},''
  \href{http://dx.doi.org/10.1103/PhysRevLett.98.021101}{{\em Phys. Rev. Lett.}
  {\bfseries 98} (2007) 021101},
\href{http://arxiv.org/abs/hep-ph/0611184}{{\ttfamily arXiv:hep-ph/0611184
  [hep-ph]}}.

\bibitem{Decca:2005qz}
R.~S. Decca, D.~Lopez, H.~B. Chan, E.~Fischbach, D.~E. Krause, and C.~R.
  Jamell, ``{Constraining new forces in the Casimir regime using the
  isoelectronic technique},''
  \href{http://dx.doi.org/10.1103/PhysRevLett.94.240401}{{\em Phys. Rev. Lett.}
  {\bfseries 94} (2005) 240401},
\href{http://arxiv.org/abs/hep-ph/0502025}{{\ttfamily arXiv:hep-ph/0502025
  [hep-ph]}}.

\bibitem{Chen:2014oda}
Y.~J. Chen, W.~K. Tham, D.~E. Krause, D.~Lopez, E.~Fischbach, and R.~S. Decca,
  ``{Stronger Limits on Hypothetical Yukawa Interactions in the 30–8000 nm
  Range},'' \href{http://dx.doi.org/10.1103/PhysRevLett.116.221102}{{\em Phys.
  Rev. Lett.} {\bfseries 116} no.~22, (2016) 221102},
\href{http://arxiv.org/abs/1410.7267}{{\ttfamily arXiv:1410.7267 [hep-ex]}}.

\bibitem{Chiaverini:2002cb}
J.~Chiaverini, S.~J. Smullin, A.~A. Geraci, D.~M. Weld, and A.~Kapitulnik,
  ``{New experimental constraints on nonNewtonian forces below 100 microns},''
  \href{http://dx.doi.org/10.1103/PhysRevLett.90.151101}{{\em Phys. Rev. Lett.}
  {\bfseries 90} (2003) 151101},
\href{http://arxiv.org/abs/hep-ph/0209325}{{\ttfamily arXiv:hep-ph/0209325
  [hep-ph]}}.

\bibitem{Geraci:2008hb}
A.~A. Geraci, S.~J. Smullin, D.~M. Weld, J.~Chiaverini, and A.~Kapitulnik,
  ``{Improved constraints on non-Newtonian forces at 10 microns},''
  \href{http://dx.doi.org/10.1103/PhysRevD.78.022002}{{\em Phys. Rev.}
  {\bfseries D78} (2008) 022002},
\href{http://arxiv.org/abs/0802.2350}{{\ttfamily arXiv:0802.2350 [hep-ex]}}.

\bibitem{Long:2002wn}
J.~C. Long, H.~W. Chan, A.~B. Churnside, E.~A. Gulbis, M.~C.~M. Varney, and
  J.~C. Price, ``{Upper limits to submillimeter-range forces from extra
  space-time dimensions},''
  \href{http://arxiv.org/abs/hep-ph/0210004}{{\ttfamily arXiv:hep-ph/0210004
  [hep-ph]}}.
[Nature421,922(2003)].

\bibitem{Tu:2007zz}
L.-C. Tu, S.-G. Guan, J.~Luo, C.-G. Shao, and L.-X. Liu, ``{Null Test of
  Newtonian Inverse-Square Law at Submillimeter Range with a Dual-Modulation
  Torsion Pendulum},''
\href{http://dx.doi.org/10.1103/PhysRevLett.98.201101}{{\em Phys. Rev. Lett.}
  {\bfseries 98} (2007) 201101}.

\bibitem{Yang:2012zzb}
S.-Q. Yang, B.-F. Zhan, Q.-L. Wang, C.-G. Shao, L.-C. Tu, W.-H. Tan, and
  J.~Luo, ``{Test of the Gravitational Inverse Square Law at Millimeter
  Ranges},''
\href{http://dx.doi.org/10.1103/PhysRevLett.108.081101}{{\em Phys. Rev. Lett.}
  {\bfseries 108} (2012) 081101}.

\bibitem{Tan:2016vwu}
W.-H. Tan, S.-Q. Yang, C.-G. Shao, J.~Li, A.-B. Du, B.-F. Zhan, Q.-L. Wang,
  P.-S. Luo, L.-C. Tu, and J.~Luo, ``{New Test of the Gravitational
  Inverse-Square Law at the Submillimeter Range with Dual Modulation and
  Compensation},''
\href{http://dx.doi.org/10.1103/PhysRevLett.116.131101}{{\em Phys. Rev. Lett.}
  {\bfseries 116} no.~13, (2016) 131101}.

\bibitem{Lee:2020zjt}
J.~G. Lee, E.~G. Adelberger, T.~S. Cook, S.~M. Fleischer, and B.~R. Heckel,
  ``{New Test of the Gravitational $1/r^2$ Law at Separations down to 52
  $\mu$m},'' \href{http://dx.doi.org/10.1103/PhysRevLett.124.101101}{{\em Phys.
  Rev. Lett.} {\bfseries 124} no.~10, (2020) 101101},
\href{http://arxiv.org/abs/2002.11761}{{\ttfamily arXiv:2002.11761 [hep-ex]}}.

\end{thebibliography}\endgroup

\end{document}